\documentclass[apj]{emulateapj}
\usepackage{graphicx}
\usepackage{txfonts}
\usepackage{float}
\usepackage{booktabs}
\usepackage{natbib}
\bibpunct{(}{)}{;}{a}{}{,} 
\usepackage{amssymb}
\usepackage{pifont}
\usepackage{ gensymb }
\usepackage{longtable}
\usepackage{url}

\shorttitle{LLAGN physics through variability}
\shortauthors{Hern\'{a}ndez-Garc\'{i}a et al.}

\begin{document}

\title{Unveiling the physics of low luminosity AGN through X-ray variability: \\
 LINER versus Seyfert 2}

\author{Hern\'{a}ndez-Garc\'{i}a, L.\altaffilmark{1}; Masegosa, J.\altaffilmark{1}; Gonz\'{a}lez-Mart\'{i}n, O.\altaffilmark{2}; M\'{a}rquez, I. \altaffilmark{1}; Perea, J. \altaffilmark{1}}

\affil{$^{1}$Instituto de Astrof\'{i}sica de Andaluc\'{i}a, CSIC, Glorieta de la Astronom\'{i}a, s/n, 18008 Granada, Spain\\
             $^{2}$Instituto de radioastronom\'{i}a y Astrof\'{i}sica (IRyA-UNAM), 3-72 (Xangari), 8701, Morelia, Mexico \\}


\begin{abstract}
X-ray variability is very common in active galactic nuclei (AGN), but these variations may not occur similarly in different families of AGN. We aim to disentangle the structure of low ionization nuclear emission line regions (LINERs) compared to Seyfert 2s by the study of their spectral properties and X-ray variations. We assembled the X-ray spectral parameters and variability patterns, which were obtained from simultaneous spectral fittings. Major differences are observed in the X-ray luminosities, and the Eddington ratios, which are higher in Seyfert 2s. 
Short-term X-ray variations were not detected, while long-term changes are common in LINERs and Seyfert 2s. Compton-thick sources generally do not show variations, most probably because the AGN is not accesible in the 0.5--10 keV energy band. The changes are mostly related with variations in the nuclear continuum, but other patterns of variability show that variations in the absorbers and at soft energies can be present in a few cases. We conclude that the X-ray variations may occur similarly in LINERs and Seyfert 2s, i.e., they are related to the nuclear continuum, although they might have different accretion mechanisms. Variations at UV frequencies are detected in LINER nuclei but not in Seyfert 2s. This is suggestive of at least some LINERs having an unobstructed view of the inner disc where the UV emission might take place, being UV variations common in them. This result might be compatible with the disappeareance of the torus and/or the broad line region in at least some LINERs.
\end{abstract}



\section{Introduction}

\begin{table*}
\begin{center}
\caption{\label{properties} General properties of the sample galaxies.}
\begin{tabular}{lcccccc} \hline
\hline
Name    & RA & DEC &  Dist.$^a$  &   Morph.  & Optical &  Variability \\
 & (J2000)  & (J2000) & (Mpc) &   type & class. &    pattern \\
(1) & (2) & (3) & (4) & (5) & (6) & (7)    \\  \hline
NGC~315   &   00 57 48.9 & +00 21 09  & 59.60 &  E &		 L1.9 &  - \\
NGC~1052  &    02 41 04.8 & +08 15 21  & 19.48 &   E & 	 L1.9 &  $Norm_2$ and $N_{H2}$ \\
NGC~1961  &  05 42 04.6 & +69 22 42	&  56.20 &     SAB(rs)c &  L2 &  - \\
NGC~2681*  &    08 53 32.7 & +51 18 49  &  15.25 &  S0-a(s) &	 L1.9   &  - \\
NGC~3718  &   11 32 34.8 & +53 04 05	 & 17.00 &   SB(s)a &  L1.9 &  $Norm_2$ \\
NGC~4261  &    12 19 23.2 & +05 49 31  & 31.32 &    E & 	 L2 &  - \\
NGC~4278  &   12 20 06.8 & +29 16 51  & 15.83 &   E &  	 L1.9 &  $Norm_2$ \\
NGC~4374*  &   12 25 03.7 & +12 53 13  & 17.18 &    E &  	 L2   &  $Norm_2$ \\
NGC~4494  &   12 31 24.0 & +25 46 30  & 13.84 &    E &  	 L2:: &  $Norm_2$ \\
NGC\,4552 & 12 35 39.8 & +12 33 23 & 15.35 &  E & L2 &  $Norm_2$ and $Norm_1$ \\
NGC~4736  &   12 50 53.1 & +41 07 14  & 5.02 &    Sab(r) &	 L2 &  - \\
NGC~5195  &   13 29 59.6 & +47 15 58	 &  7.91 &   IA &  	 L2: &  $Norm_2$ \\
NGC~5982  &   15 38 39.8 & +59 21 21	 & 41.22 &   E &		 L2:: &  $Norm_2$ \\ 
\hline
MARK\,348   &       00 48 47.2 &   +31 57 25 & 63.90  &    S0-a  &  S2 &   $Norm_2$ \\ 
NGC\,424*    &       01 11 27.7 &  -38  05  01 &  47.60 &    S0-a  &  S2  &  - \\ 
MARK\,573*   &       01 43 57.8 &    +02 20 59 & 71.30  &    S0-a   &  S2  &  - \\ 
NGC\,788    &       02  01  06.5 &  -06 48 56 & 56.10  &    S0-a   &  S2 &  -  \\ 
 ESO\,417-G06   &    02 56 21.5 &  -32 11  06 & 65.60 &     S0-a   & S2 &  $N_{H2}$ \\ 
 MARK\,1066*     &    02 59 58.6 &   +36 49 14 & 51.70 &     S0-a  &  S2  &  - \\  
 3C\,98.0      &    03 58 54.5 &   +10 26  02 & 124.90 &     E  & S2 &  $Norm_2$ \\  
MARK\,3*     &       06 15 36.3 &   +71  02 15 & 63.20  &    S0   &  S2  &  $Norm_2$ \\ 
 MARK\,1210     &    08  04  05.9 &    +05  06 50 & 53.60 &     -  & S2 &  $Norm_2$ and $N_{H2}$ \\   
 IC\,2560*       &   10 16 19.3 &  -33 33 59 & 34.80 &     SBb  & S2  &  - \\  
 NGC\,3393*      &   10 48 23.4 &  -25  09 44   & 48.70 &    SBa   &  S2   &  - \\ 
NGC\,4507   &      12 35 36.5 &  -39 54 33 &  46.00 &    Sab    &  S2 &  $Norm_2$ and $N_{H2}$ \\  
 NGC\,4698      &   12 48 22.9 &    +08 29 14 & 23.40 &     Sab   & S2 &  - \\  
 NGC\,5194*      &   13 29 52.4 &   +47 11 41 & 7.85 &    Sbc   & S2  &  - \\   
 MARK\,268     &   13 41 11.1 &   +30 22 41 & 161.50 &     S0-a   & S2 &  - \\  
 MARK\,273     &   13 44 42.1 &   +55 53 13 & 156.70 &     Sab   & S2 &  $N_{H2}$ \\  
Circinus*    &      14 13  09.8 &  -65 20 17 & 4.21  &     Sb   &  S2  &  - \\
 NGC\,5643*      &   14 32 40.7 &  -44 10 28 & 16.90 &     Sc  & S2  &  - \\  
 MARK\,477*     &   14 40 38.1 &   +53 30 15 & 156.70 &    E?    & S2   &  - \\  
 IC\,4518A      &   14 57 41.2 &  -43  07 56 & 65.20 &     Sc  & S2 &  $Norm_2$ \\   
 ESO\,138-G01*   &   16 51 20.5 &  -59 14 11 & 36.00 &     E-S0  & S2   &  - \\  
 NGC\,6300      &   17 16 59.2 &  -62 49  05 & 14.43 &     SBb   & S2 &  $Norm_2$ and $Norm_1$ \\  
 NGC\,7172      &   22  02  01.9 &  -31 52  08 & 33.90 &     Sa  & S2 &  $Norm_2$ \\  
NGC\,7212*   &      22  07  02.0 &   +10 14  00 & 111.80  &     Sb  &  S2   &  - \\
 NGC\,7319      &   22 36  03.5 &   +33 58 33 & 77.25 &    Sbc    & S2 &  $Norm_2$ and $N_{H1}$ \\ 
 \hline
\end{tabular} \\
\end{center}
{ {\bf Notes.} (Col. 1) Name (those marked with asterisks are Compton-thick candidates), (Col. 2) right ascension, (Col. 3) declination, (Col. 4) distance, (Col. 5)  galaxy morphological type from \cite{omaira2009b} or Hyperleda, (Col. 6) optical classification, where L: LINER (quality ratings as described by \cite{ho1997} are given by ``:'' and ``::'' for uncertain and highly uncertain classification, respectively) and S2: Seyfert 2, and (Col. 7) X-ray variability pattern, where the parameters that vary in the model refer to the normalizations at soft ($Norm_1$) and hard ($Norm_2$) energies, and/or the absorber at soft ($N_{H1}$) and hard energies ($N_{H2}$), and the lines mean that variations are not detected. $^a$All distances are taken from the NED and correspond to the average redshift-independent distance estimates, when available, or to the redshift-estimated distance otherwise.} \\
\vspace*{-0.5cm}

\end{table*}

Active galactic nuclei (AGN) include a number of subgroups that
are thought to be represented under the same scenario, the unified model (UM) of AGN \citep{antonucci1993}. Under
this scheme, the differences between objects are attributed only
to orientation effects. 
When this picture was drawn, some subsamples of AGN were not taken into account, as it is the case of low ionization nuclear emission line regions (LINERs), and in fact they do not fit within the UM \citep{ho2008}.
It has been assumed that LINERs are scaled down versions of more powerful AGN (i.e., Seyferts and quasars) but, as noted by \cite{ho2008}, a key point is that the picture is not so simple, since slight differences between LINERs and Seyferts have been found at different wavelengths.
Recent observations indeed suggest that
the UM should be slightly modified \citep[see][for a recent revision and also \citealt{antonucci2013}]{netzer2015}, including the nature of
the torus, which some authors suggest might be clumpy \citep[e.g.,][]{nenkova2008, stalevski2012} and can disappear at low luminosities \citep[e.g.,][]{elitzur2006}. On the other hand, recent works suggest also a dependence of accretion state on luminosity, black hole mass, and galaxy evolution, in the sense of being more efficient for less massive supermassive black holes (SMBH) and more luminous AGN \citep[e.g.,][]{gucao2009,schawinski2012,yang2015}.

X-ray energies provide the best way to study the physical mechanism operating in AGN since they 
have the power of penetrating
through the dusty torus so the inner parts of the AGN can
be accessed \citep{awaki1991,turner1997,maiolino1998}. Moreover, strong and random variability on a wide range of timescales and
wavelengths can be considered the best evidence of an AGN. 
Its study is a highly valuable tool for the comprehension of the physical structure of AGN. 
In particular, the X-ray flux exhibits variability
on time scales shorter than any other energy band \citep[e.g.,][]{vaughan2003,mchardy2013}, indicating that the emission occurs in the innermost regions
of the central engine. Therefore, X-ray variability provides a powerful tool to probe the extreme physical
processes operating in the inner parts of the accretion flow close to the SMBH.
X-ray variability
has been found in almost all AGN analyzed families, from the highest luminosity
regime, i.e., quasars \citep{schmidt1963, mateos2007}, through Seyferts \citep[][hereinafter HG+15]{risaliti2000,evans2005,panessa2011, risaliti2011, lore2015}, to the lowest luminosity regime, e.g., LINERs \citep[][hereinafter HG+13 and HG+14]{pian2010,younes2011, lore2013, lore2014}. However, it is still under debate
what is the mechanism responsible for those variations, as
well as whether the changes occur similarly in every AGN. 

In previous works, we have studied the X-ray spectral variability of two subgroups of AGN, selected from their optical classifications as LINERs (HG+13 and HG+14) and Seyfert 2 galaxies (HG+15). 
The data were obtained from the public archives of \emph{Chandra} and/or \emph{XMM}--Newton, and the same method was used to search for their variability pattern(s) in both subgroups. 
In this work we present the X-ray spectral properties derived from these analyses, as well as the X-ray variability pattern(s) obtained for LINER and Seyfert 2 galaxies, with the aim of finding similarities and/or differences within the two families of AGN. This is important to understand how is the internal structure of LINERs compared to Seyfert 2s, their closest (in properties) AGN family, as shown by \cite{omaira2014} using artificial neural networks.
 
This paper is organised as follows: the sample used for the work is described in Sect. 2, the methodology used to derive the physical parameters and the variability patterns is described in Sect. 3, the results of the comparison of the X-ray variability and spectral properties between the two families is presented in Sect. 4, which are discussed in Sect. 5. A summary of the main results is given in Sect. 6.


\section[]{\label{sample}Sample and data}

Our sample was selected for having more than one observations in the public archives of the X-ray satellites \emph{Chandra} and/or \emph{XMM}--Newton.
The sample contains 21 LINERs, 18 from the Palomar sample \citep{ho1997} and seven from the sample included in \cite{omaira2009a} -- with four sources in common -- and 26 Seyfert 2s from the V\'{e}ron-Cetty and V\'{e}ron catalogue \citep{veron2010}. The data used for this work is presented in HG+13, and HG+14 for LINER nuclei, and in HG+15 for Seyfert 2s. Thus we refer the reader to these papers for details on the sample selection.
Since most LINERs were selected from \cite{ho1997}, who followed the diagnostic diagrams proposed by \cite{veilleuxosterbrock1987} for classification purposes, we verified that all the sources are consistent with these criteria. For LINERs, \cite{ho1997} and \cite{omaira2009a} used the same diagnostic diagrams for their classifications. For the classification of Seyfert 2s we verified that all of them are consistent with the criteria in \cite{veilleuxosterbrock1987} -- the references can be found in Appendix B from HG+15.

Some of the galaxies have been rejected for the present analysis: the LINERs NGC\,2787, NGC\,2841, and NGC\,3627 and the Seyfert 2 NGC\,3079 due to the strong extranuclear emission contamination; the LINER NGC\,3226 due to its contamination from the companion galaxy NGC\,3227; and all the LINERs classified as non-AGN by \citet[][NGC\,3608, NGC\,4636, NGC\,5813, and NGC\,5846]{omaira2009b} due to the lack of analog sources among Seyferts.

All together, the sample contains a total of 38 sources: 13 LINERs (two Compton-thick candidates\footnote{We considered a Compton-thick candidate (i.e., $N_H > 1.5 \times 10^{24} cm^ {-2}$) when at least two of the following criteria were met: $\Gamma<1$, $EW(Fek\alpha)>500eV$, and $F(2-10 keV)/F([OIII])<1$.} and 11 Compton-thin\footnote{Classifications are obtained from \cite{omaira2009b}. Four sources were not included in their sample and are Compton-thin (NGC\,1961, NGC\,3718, NGC\,5195, and NGC\,5982) based on the values of $\Gamma$ and the X-ray to [O III] flux ratio (see \citealt{omaira2009b}).}) and 25 Seyfert 2s (12 Compton-thick candidates and 13 Compton-thin). 
Compton-thick LINERs are not considered as a group due to their low numbers, and will not be discussed further. This has no impact on the derived results (see Sect. 4).
Table \ref{properties} shows the list of the sample galaxies, along with its X-ray variability pattern (Col. 7). These variability patterns are related to the normalizations at soft ($Norm_1$) and hard ($Norm_2$) energies, and/or the absorber at soft ($N_{H1}$) and hard energies ($N_{H2}$, see HG+13 for more details). We refer the reader to Sect. \ref{simultaneous} for details on the variability pattern.

In order to test whether we are able to compare the spectra of LINERs and Seyfert 2s, we calculated the signal-to-noise ratio (S/N) of each spectrum individually in the 0.5--2 keV and 2--10 keV energy bands following the formulae given in \cite{stoehr2008}. We found that the S/N cover the same range of values in both families and is independent of the spectral modeling (i.e., spectra with larger S/N does not necessarily require more complex models), concluding that the spectra are comparable. The median [25\% and 75\% quartiles] are S/N(0.5--2 keV) = 6.33[5.23--7.23] and S/N(2--10 keV) = 5.36[4.07--6.51] for LINERs and S/N(0.5--2 keV) = 5.17[3.75--6.39] and S/N(2--10 keV) = 3.84[2.32--5.19] for Seyfert 2s.


\section{\label{method}Methodology}

In this section, we will briefly summarize the methodology used to obtain the X-ray spectral parameters and variability patterns of the sources. This information has been taken from HG+13, HG+14 and HG+15, thus we refer the reader to those papers for more details.

\subsection{X-ray short-term variablity}

X-ray variations between hours and days were studied from the analysis of the light curves. We calculated the normalized excess
variance, $\rm{\sigma_{NXS}^2}$, for each light curve segment with
30-40 ksec following prescriptions in \cite{vaughan2003} \citep[see
  also][]{omaira2011a}. We considered short-term variations for $\rm{\sigma_{NXS}^2}$
detections above 3$\sigma$ of the confidence level.

\subsection{X-ray long-term variability}

X-ray spectral variations between weeks and years were studied in two steps. Firstly, we analyzed each observation separately, and secondly we analyzed together all the spectra of the same nucleus.

\subsubsection{\label{indiv}Individual spectral analysis}

We first performed a spectral fit to each observation individually using the following models:

\begin{itemize}
\item[$\bullet$] 
\underline{PL:} $e^{N_{Gal} \sigma (E)} \cdot e^{N_{H} \sigma (E(1+z))}[N_{H}] \cdot Norm e^{-\Gamma}[\Gamma, Norm].$
\vspace*{0.2cm}
\item[$\bullet$] \underline{ME:} 
$e^{N_{Gal} \sigma (E)} \cdot e^{N_{H} \sigma (E(1+z))}[N_{H}] \cdot MEKAL[kT, Norm].$
\vspace*{0.2cm}

\item[$\bullet$] \underline{2PL: }
$e^{N_{Gal} \sigma (E)} \big( e^{N_{H1} \sigma (E(1+z))}[N_{H1}] \cdot Norm_1 e^{-\Gamma}[\Gamma, Norm_1] + e^{N_{H2} \sigma (E(1+z))}[N_{H2}] \cdot Norm_2 e^{-\Gamma}[\Gamma, Norm_2]\big)$.
\vspace*{0.2cm}

\item[$\bullet$] \underline{MEPL: }
$e^{N_{Gal} \sigma (E)} \big(e^{N_{H1} \sigma (E(1+z))}[N_{H1}] \cdot MEKAL[kT, Norm_1] + e^{N_{H2} \sigma (E(1+z))}[N_{H2}] \cdot Norm_2 e^{-\Gamma}[\Gamma, Norm_2]\big)$.
\vspace*{0.2cm}

\item[$\bullet$] \underline{ME2PL:}
$e^{N_{Gal} \sigma (E)} \big( e^{N_{H1} \sigma (E(1+z))}[N_{H1}] \cdot Norm_1 e^{-\Gamma}[\Gamma, Norm_1] + MEKAL[kT] + e^{N_{H2} \sigma (E(1+z))}[N_{H2}] \cdot Norm_2 e^{-\Gamma}[\Gamma, Norm_2]\big)$.
\vspace*{0.2cm}

\item[$\bullet$] \underline{2ME2PL:} 
$e^{N_{Gal} \sigma (E)} \big( e^{N_{H1} \sigma (E(1+z))}[N_{H1}] \cdot Norm_1 e^{-\Gamma}[\Gamma, Norm_1] + MEKAL[kT_1] + MEKAL[kT_2] + e^{N_{H2} \sigma (E(1+z))}[N_{H2}] \cdot Norm_2 e^{-\Gamma}[\Gamma, Norm_2]\big)$.

\end{itemize}

\noindent In the equations above, $\sigma (E)$ is the photo-electric
cross-section, $z$ is the redshift, and $Norm_i$ are  the thermal component and the
normalizations of the power law (i.e., $Norm_1$ and $Norm_2$). For
each model, the parameters that are allowed to vary are written in brackets.  The
Galactic absoption, $N_{Gal}$, is included in each model and fixed to
the predicted value using the tool
{\sc nh} within {\sc ftools} \citep{dickeylockman1990, kalberla2005}.
All the models include three narrow Gaussian lines to take the iron lines at 6.4 keV
(FeK$\alpha$), 6.7 keV (FeXXV), and 6.95 keV (FeXXVI) into account.
The $\chi^2/d.o.f$ and F--test were used to select the simplest model
that best represents the data.

\subsubsection{\label{simultaneous}Simultaneous spectral fit}

From the individual best-fit model, and whenever the models differ for different observations\footnote{Note that the best individual fit is usually the same for observations of the same source (HG+14, HG+15).}, we chose the most complex model that fits each object to simultaneously fit all the spectra obtained at different dates of the same source. Initially, the values of the spectral parameters were set to those obtained for the spectrum with the largest number counts (which usually correspond to those with the highest S/N) for each galaxy, although note that these are included only as the initial conditions and were set free to change their values to fit all the data set. When this model resulted in a good fit of the whole data set, we considered that the source did not show spectral variations in the analyzed timescales. If this was not a good fit, we let the parameters $N_{H1}$, $N_{H2}$, $\Gamma$, $Norm_1$, $Norm_2$, $kT_1$, and $kT_2$ vary one-by-one in the model. In some cases two variable parameters were needed to obtain a good simultaneous spectral fit. In order to test whether variations in one or more parameters were needed to explain the simultaneous fit, a $\chi^2_r$ in the range between 0.9--1.4 -- and as closest to the unity as possible -- and F--test with values lower than $10^{-5}$ were used to confirm an improvement of the fit on each step. This procedure gives the variability pattern of the source.

Whenever possible, i.e., when data from the same instrument were available at different dates, we compared data from the same instrument, but we also compared \emph{Chandra} and \emph{XMM}--Newton observations. Since the extraction appertures are different for \emph{Chandra} ($\sim$ 3'') and \emph{XMM--Newton} ($\sim$ 25''), we took into account the extranuclear emission in the \emph{XMM--Newton} data by applying the models described in \ref{indiv} to the same extranuclear region in \emph{Chandra} data. This model (with its parameters frozen) was added to the \emph{XMM--Newton} model. Therefore we can compare the nuclear regions. This procedure again gives the variability pattern of the source.

Additionally, this analysis allows to obtain the spectral parameters for each source, including $N_{H1}$, $N_{H2}$, $\Gamma$, $kT_1$, and $kT_2$, and their physical properties, including the X-ray luminosities at soft, L(0.5--2 keV), and hard, L(2--10 keV), energies, and the Eddington ratios, $R_{Edd}$, which are compared in the present work for LINERs and Seyfert 2s. When a source showed variations in one of these parameters, we calculated the weighted (with the errors) mean as the value of such a parameter. It is worth noting that the $\Gamma$ used for the comparison were obtained from the individual observation with the highest S/N, as they are the most reliable values of this parameter.

\subsection{UV long-term variations}

When data from the optical monitor (OM) onboard \emph{XMM--Newton}
were available, UV luminosities were
estimated in the available filters, in simultaneous with the X-ray data.  We recall that UVW2 is centered
at 1894$\AA$ (1805-2454)$\AA$, UVM2 at 2205$\AA$ (1970-2675)$\AA$,
and UVW1 at 2675$\AA$ (2410-3565)$\AA$.  We used the OM observation
FITS source lists
(OBSMLI)\footnote{ftp://xmm2.esac.esa.int/pub/odf/data/docs/XMM-SOC-GEN-ICD-0024.pdf}
to obtain the photometry.
We assumed an object to be variable when the square root of the
squared errors was at least three times smaller than the difference
between the luminosities \citep{lore2015}.

\begin{figure*}
\centering
\includegraphics[width=0.4\textwidth]{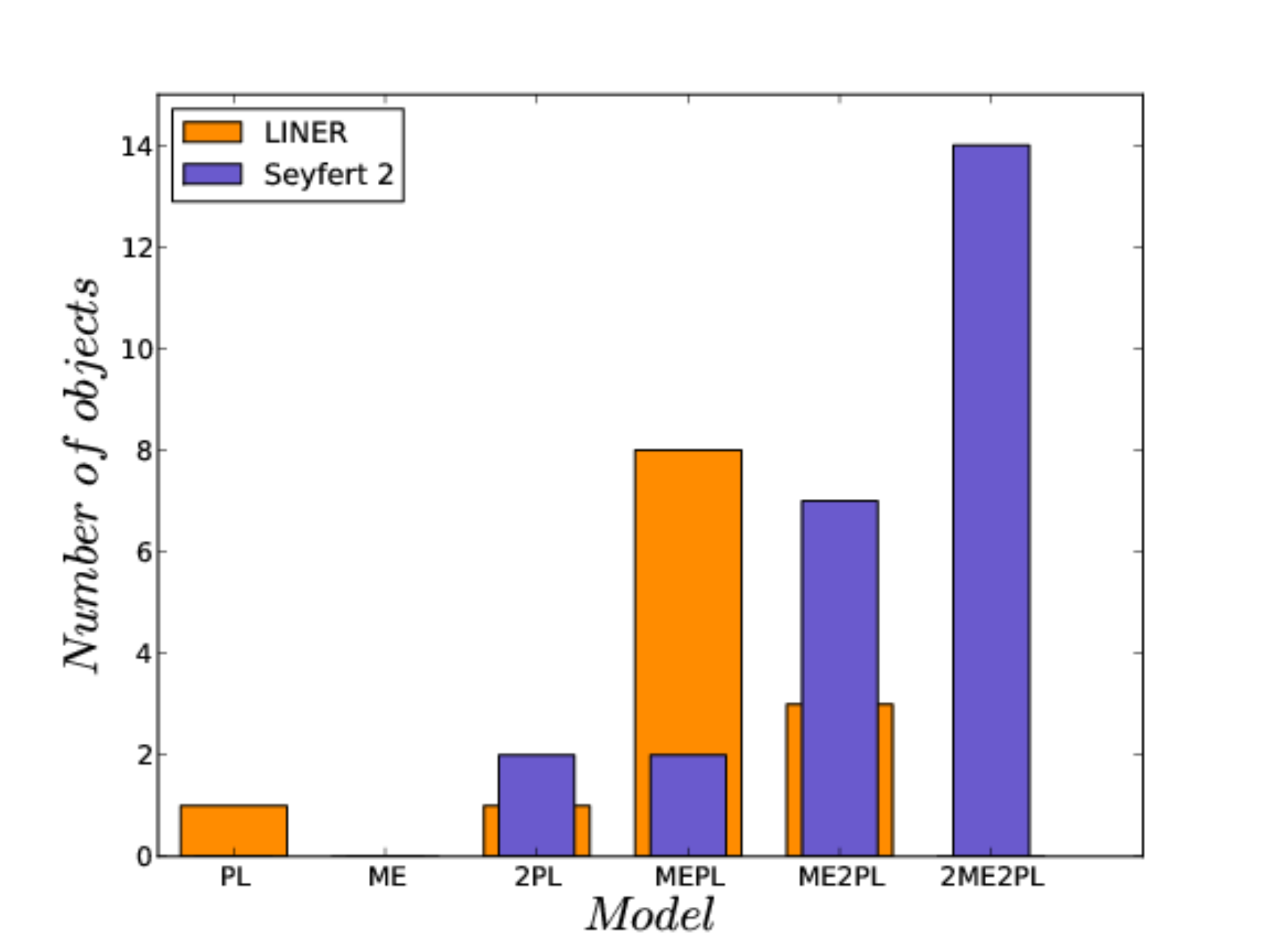}
\includegraphics[width=0.4\textwidth]{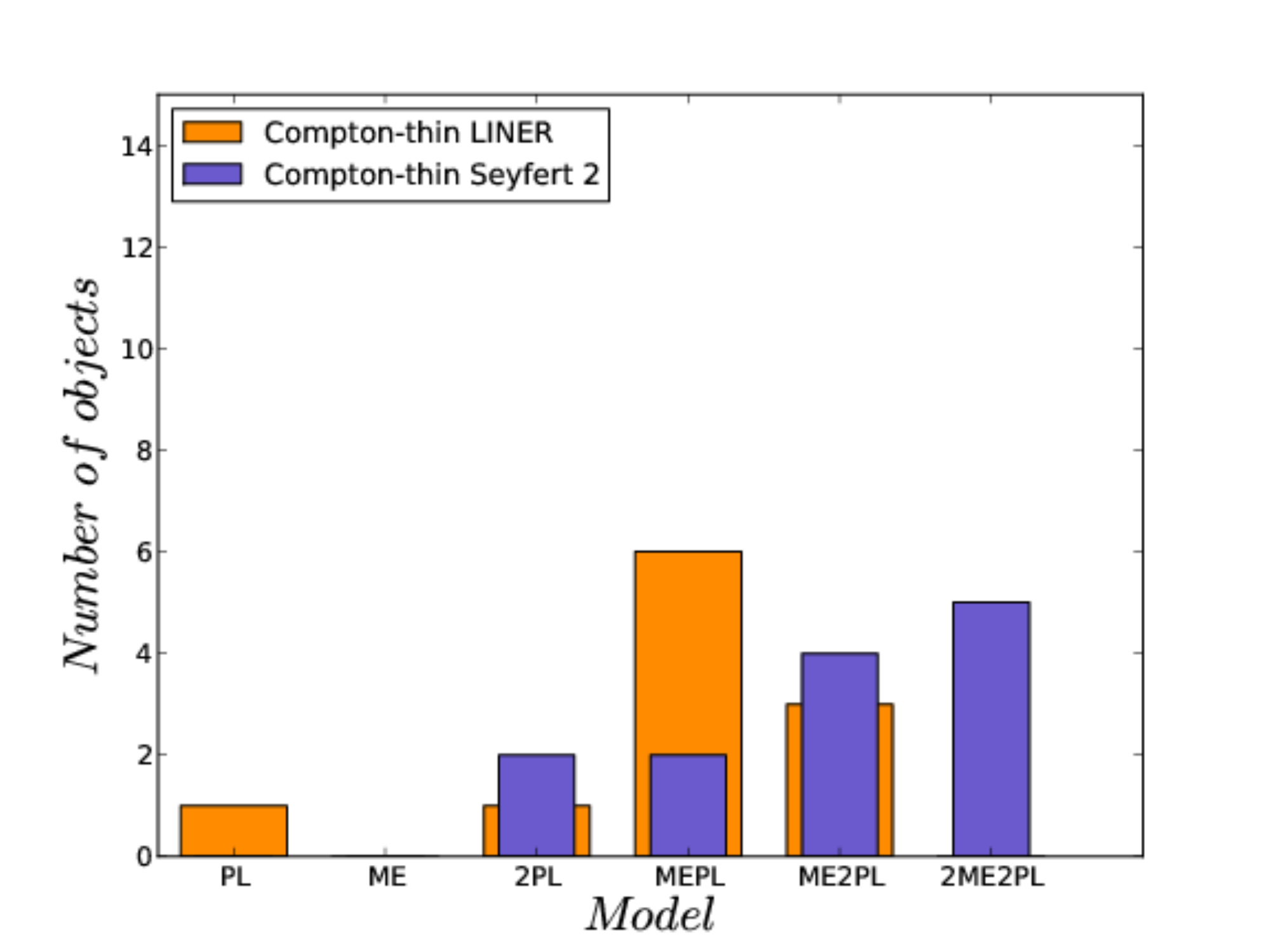}
\caption{\label{models} Histograms of the X-ray spectral models fitted to (left): all the LINERs and Seyfert 2s in the sample, and (right): Compton-thin LINERs and Seyfert 2s. Details on the models can be found in Sect. \ref{indiv}.}
\end{figure*}

\begin{table*}
\begin{center}
\footnotesize
\caption{\label{means} Median values and the 25\% and 75\% quartiles of the spectral parameters.}
\begin{tabular}{lccccc} \hline
\hline
 & \multicolumn{2}{c}{LINER} & \multicolumn{3}{c}{Seyfert 2} \\ \cmidrule(r){2-3} \cmidrule{4-6}
 & All & Compton-thin & All & Compton-thick & Compton-thin \\ 
(1) & (2) & (3) & (4) & (5) & (6)  \\ \hline \vspace*{0.1cm}
log(L(0.5-2 keV) [erg \hspace{0.1cm} $s^{-1}$]) & 39.8$_{39.5}^{41.0}$ & 40.7$_{39.5}^{41.0}$ & 41.8$_{41.3}^{42.3}$ & 41.7$_{40.6}^{42.2}$ & 42.1$_{41.3}^{42.6}$ \\ \vspace*{0.1cm}
log(L(2-10 keV) [erg \hspace{0.1cm} $s^{-1}$])   & 39.8$_{39.5}^{41.0}$ & 40.5$_{39.5}^{41.2}$ & 42.5$_{41.4}^{42.8}$ & 41.5$_{41.0}^{42.6}$ & 42.7$_{42.5}^{42.8}$ \\ \vspace*{0.1cm}
L(0.5-2 keV)/L(2-10 keV) & 0.9$_{0.6}^{1.1}$ & 0.8$_{0.6}^{1.1}$ & 0.4$_{0.3}^{0.9}$ & 0.8$_{0.3}^{1.0}$ & 0.4$_{0.3}^{0.5}$ \\ \vspace*{0.1cm}
log($M_{BH}$ [$M_{\odot}$])  & 8.4$_{7.6}^{8.7}$ & 8.4$_{7.6}^{8.7}$ & 7.5$_{7.2}^{7.8}$ & 7.4$_{6.7}^{7.8}$ & 7.6$_{7.4}^{7.8}$ \\ \vspace*{0.1cm}
log($R_{Edd}$) & -5.2$_{-5.6}^{-4.5}$ & -5.1$_{-5.6}^{-4.2}$ & -2.3$_{-2.9}^{-1.8}$ & -2.9$_{-3.0}^{-1.9}$ & -1.9$_{-2.3}^{-1.8}$ \\ \vspace*{0.1cm}
$N_{H1}$ ($\times 10^{22} [cm^{-2}]$) & 0.00$_{0.00}^{0.02}$ & 0.0$_{0.0}^{0.0}$ & 0.0$_{0.0}^{0.0}$ & 0.0$_{0.0}^{0.0}$ & 0.00$_{0.00}^{0.02}$ \\ \vspace*{0.1cm}
$N_{H2}$ ($\times 10^{22} [cm^{-2}]$) & 1.1$_{0.2}^{10.5}$ & 9.4$_{0.8}^{10.5}$ & 31.4$_{22.2}^{44.7}$ & 43.3$_{29.8}^{48.7}$ & 22.2$_{9.8}^{38.4}$ \\ \vspace*{0.1cm}
$\Gamma$ & 1.7$_{1.6}^{1.9}$ & 1.7$_{1.5}^{1.9}$ & 1.0$_{0.5}^{1.7}$ & 0.5$_{0.4}^{0.8}$ & 1.7$_{1.5}^{2.0}$ \\ \vspace*{0.1cm}
kT [keV] & 0.59$_{0.54}^{0.60}$ & 0.58$_{0.54}^{0.60}$ & 0.67$_{0.63}^{0.71}$ (0.14$_{0.11}^{0.15}$) & 0.65$_{0.61}^{0.68}$ (0.11$_{0.10}^{0.15}$) & 0.71$_{0.67}^{0.81}$(0.15$_{0.12}^{0.18}$) \\
\hline
\end{tabular} \\
\end{center}
{ {\bf Notes.} (Col. 1) Spectral parameter, (Col. 2) all the LINERs in the sample, (Col. 3) Compton-thin LINERs, (Col. 4) all the Seyfert 2s in the sample, (Col. 5) Compton-thick candidate Seyfert 2s, and (Col. 6) Compton-thin Seyfert 2s.}
\end{table*}

\begin{figure*}
\centering
{\includegraphics[width=0.4\textwidth]{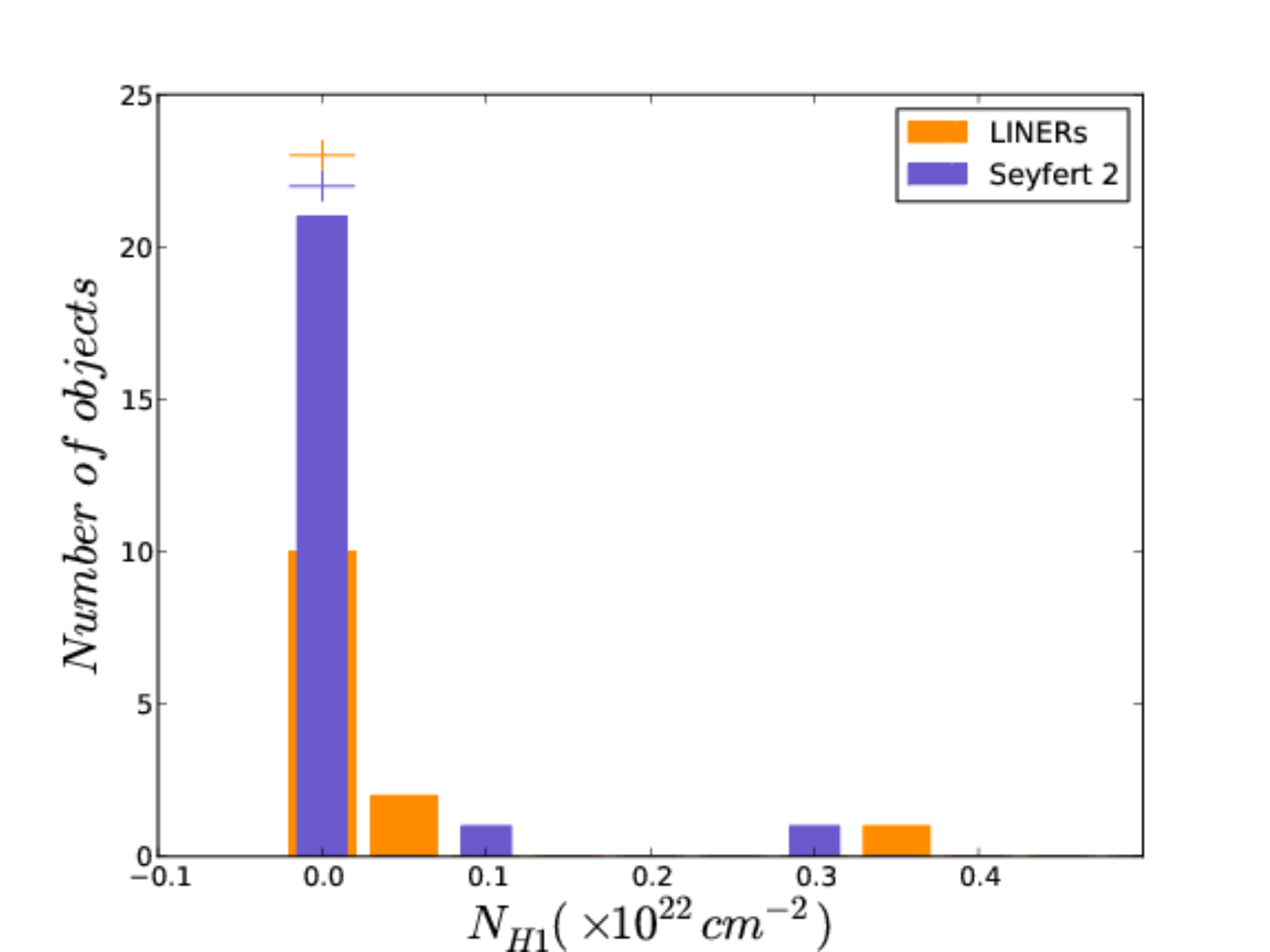}}
{\includegraphics[width=0.4\textwidth]{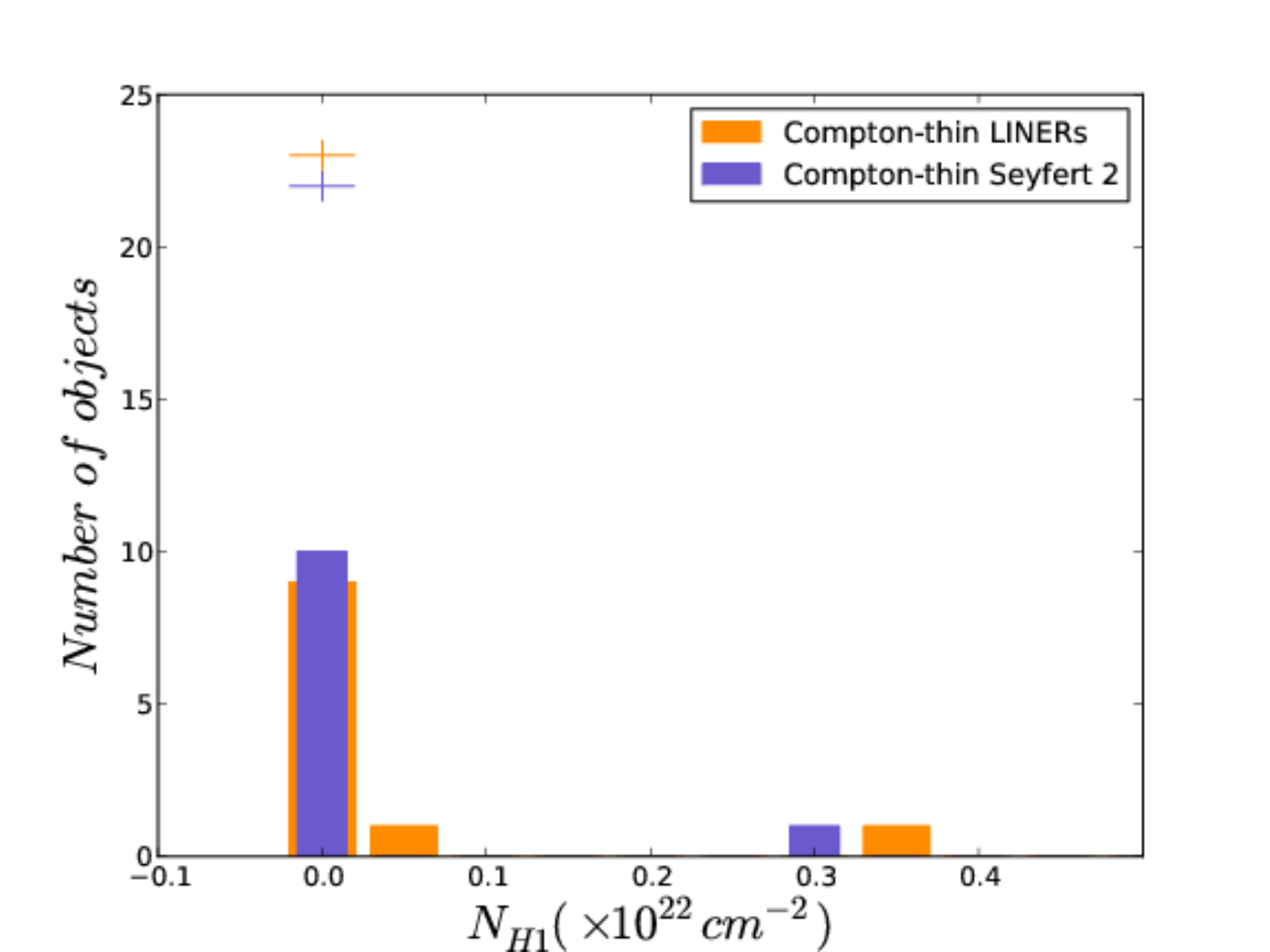}}

{\includegraphics[width=0.4\textwidth]{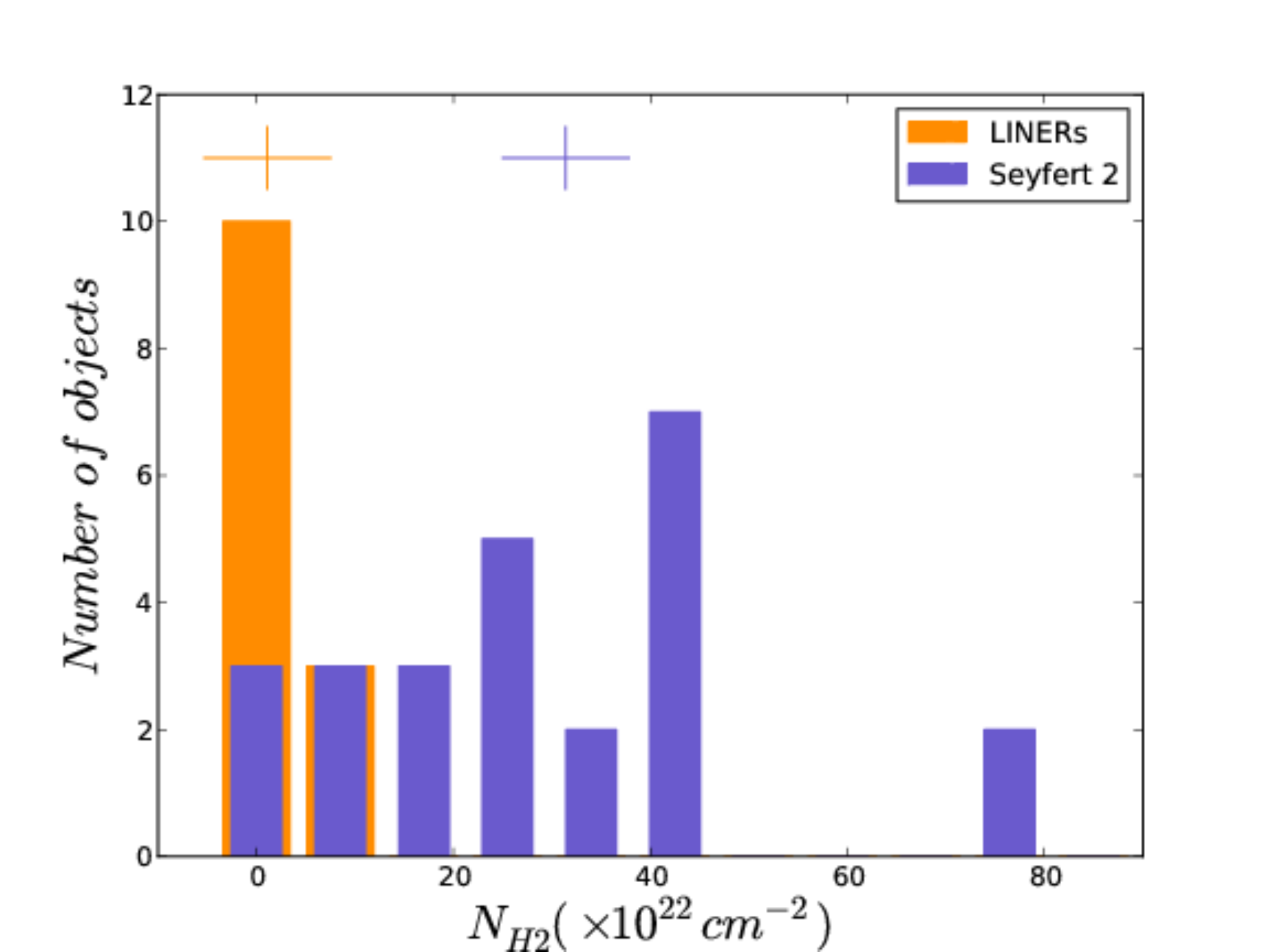}}
{\includegraphics[width=0.4\textwidth]{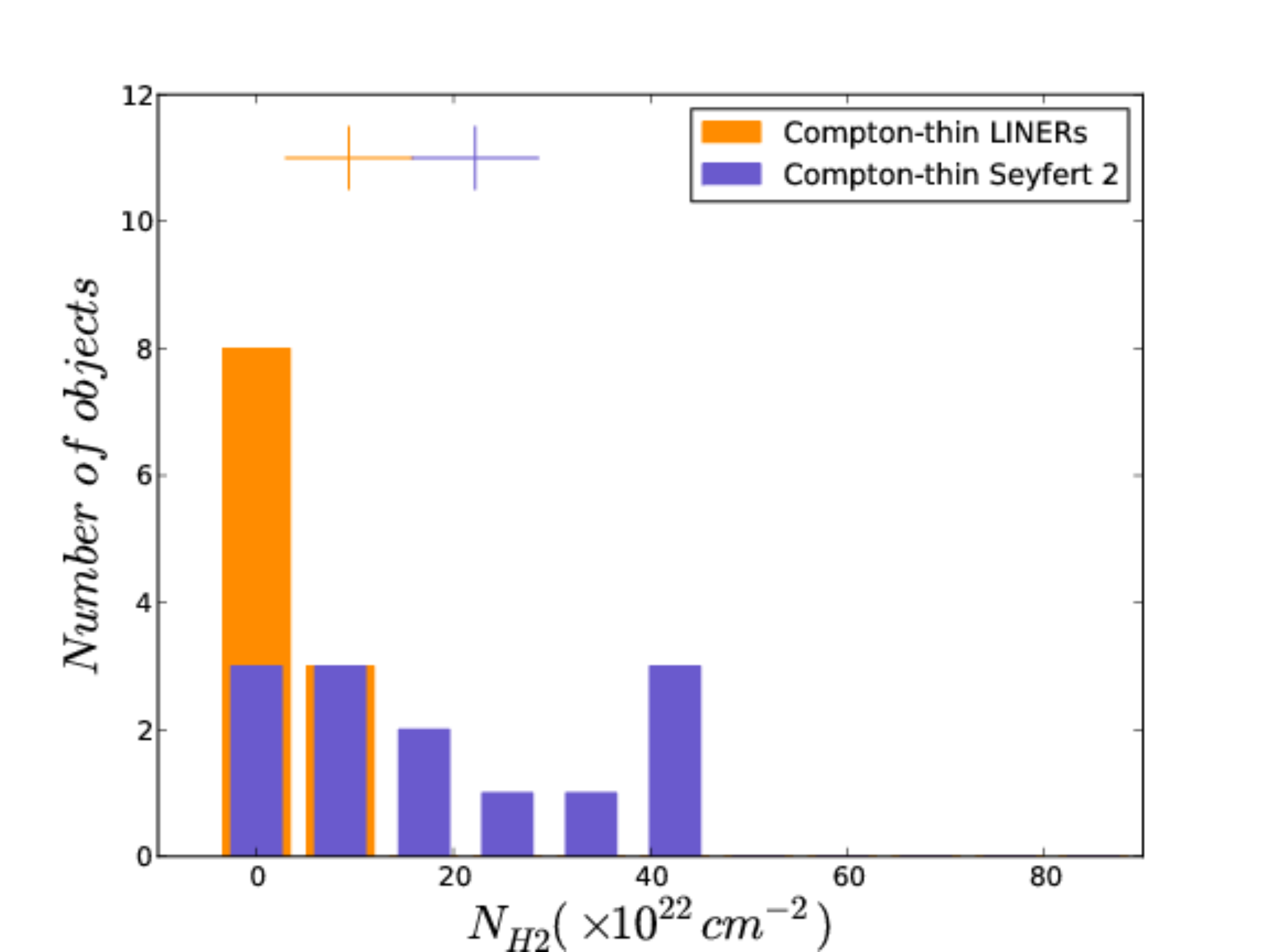}}

{\includegraphics[width=0.4\textwidth]{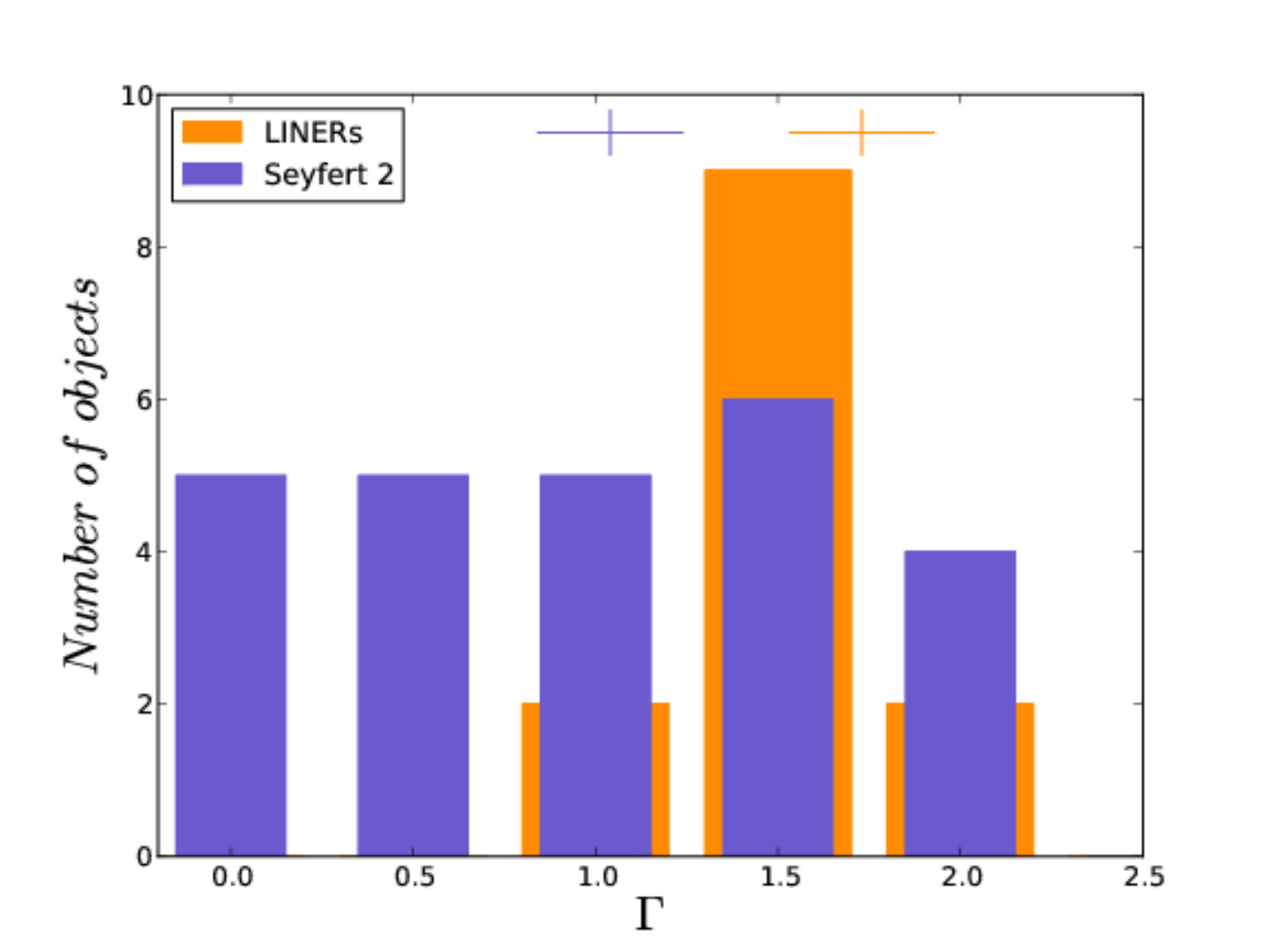}}
{\includegraphics[width=0.4\textwidth]{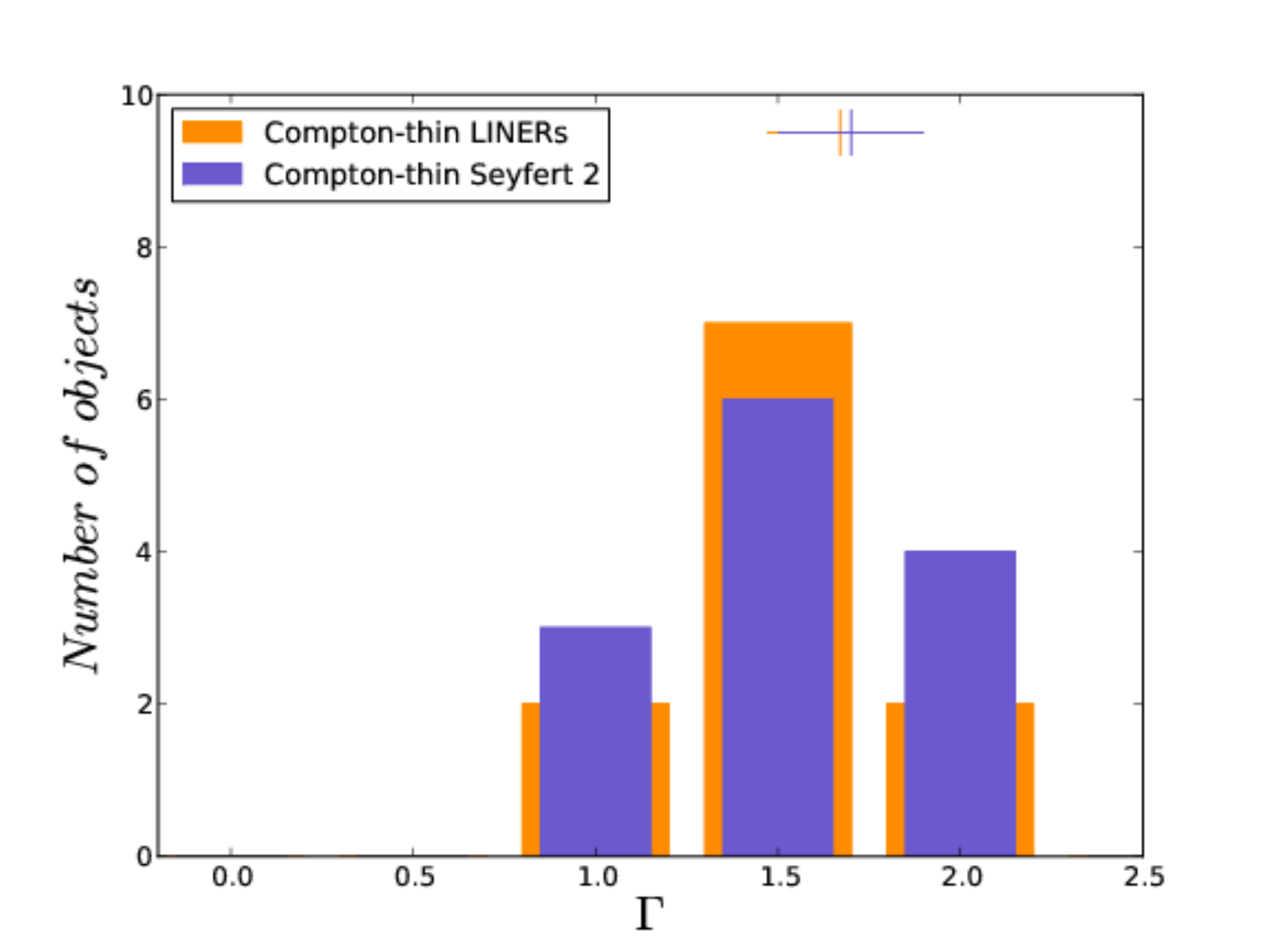}}

{\includegraphics[width=0.4\textwidth]{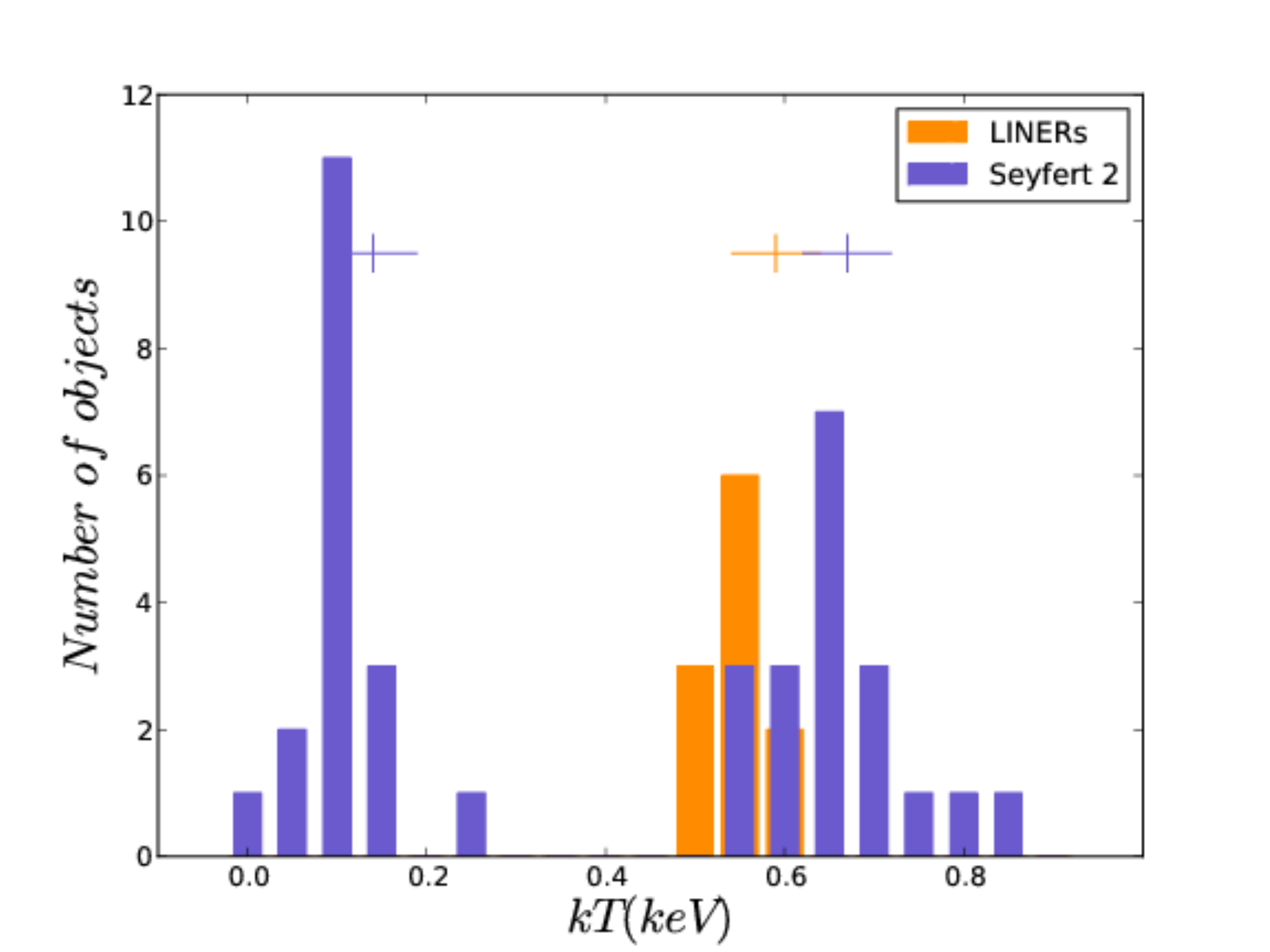}}
{\includegraphics[width=0.4\textwidth]{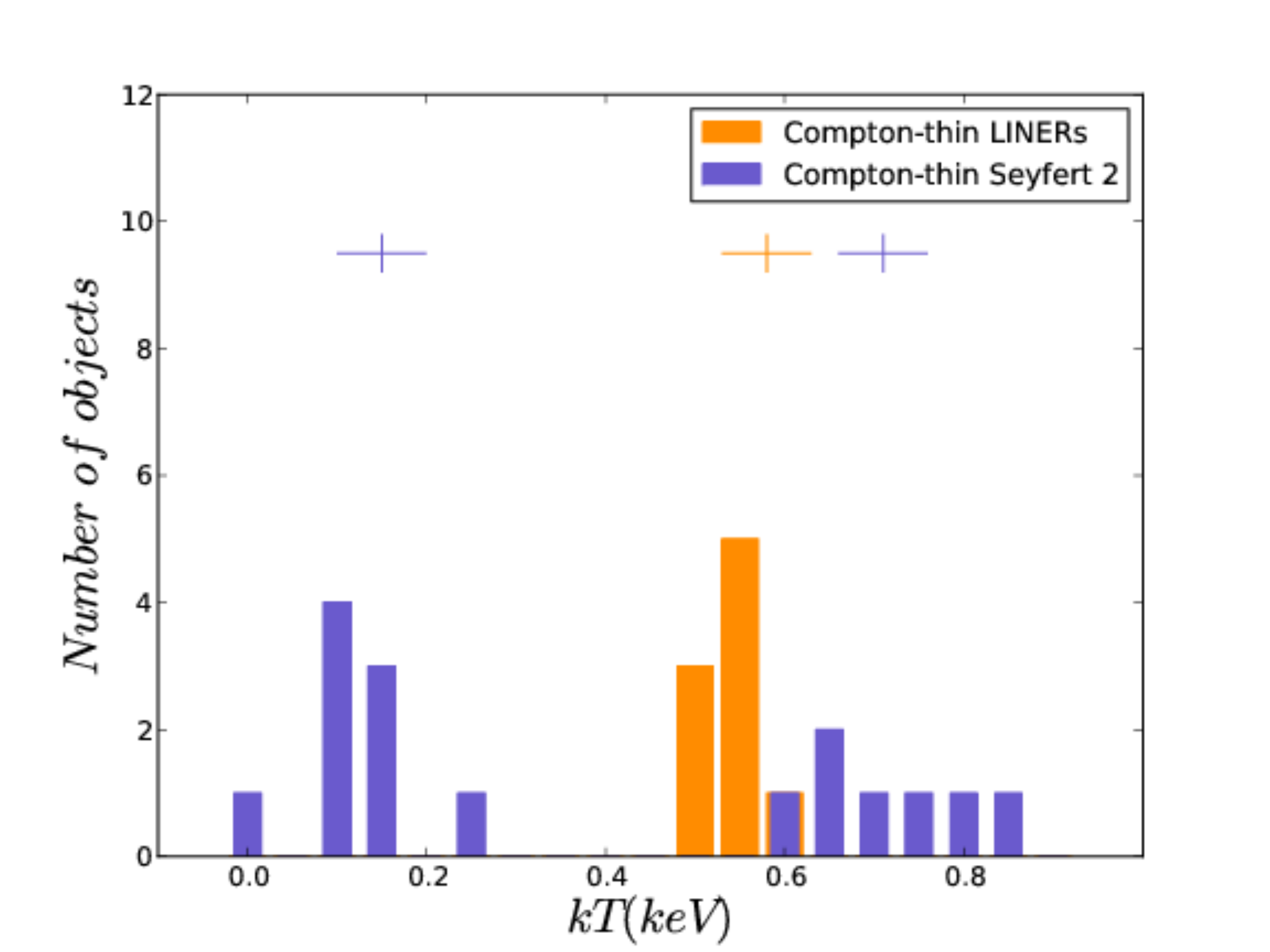}}
\caption{\label{spectral} Histograms of the main spectral parameters. From upper to lower panels: the column density at soft energies, $N_{H1}$, the column density at hard energies, $N_{H2}$, the slope of the power law, $\Gamma$, and temperatures, $kT$, are presented. In all cases (left): all the LINERs and Seyfert 2s in the sample, and (right): Compton-thin LINERs and Seyfert 2s. The crosses represent the median values reported in Table \ref{means}.}
\end{figure*}

\begin{figure*}
\centering
{\includegraphics[width=0.4\textwidth]{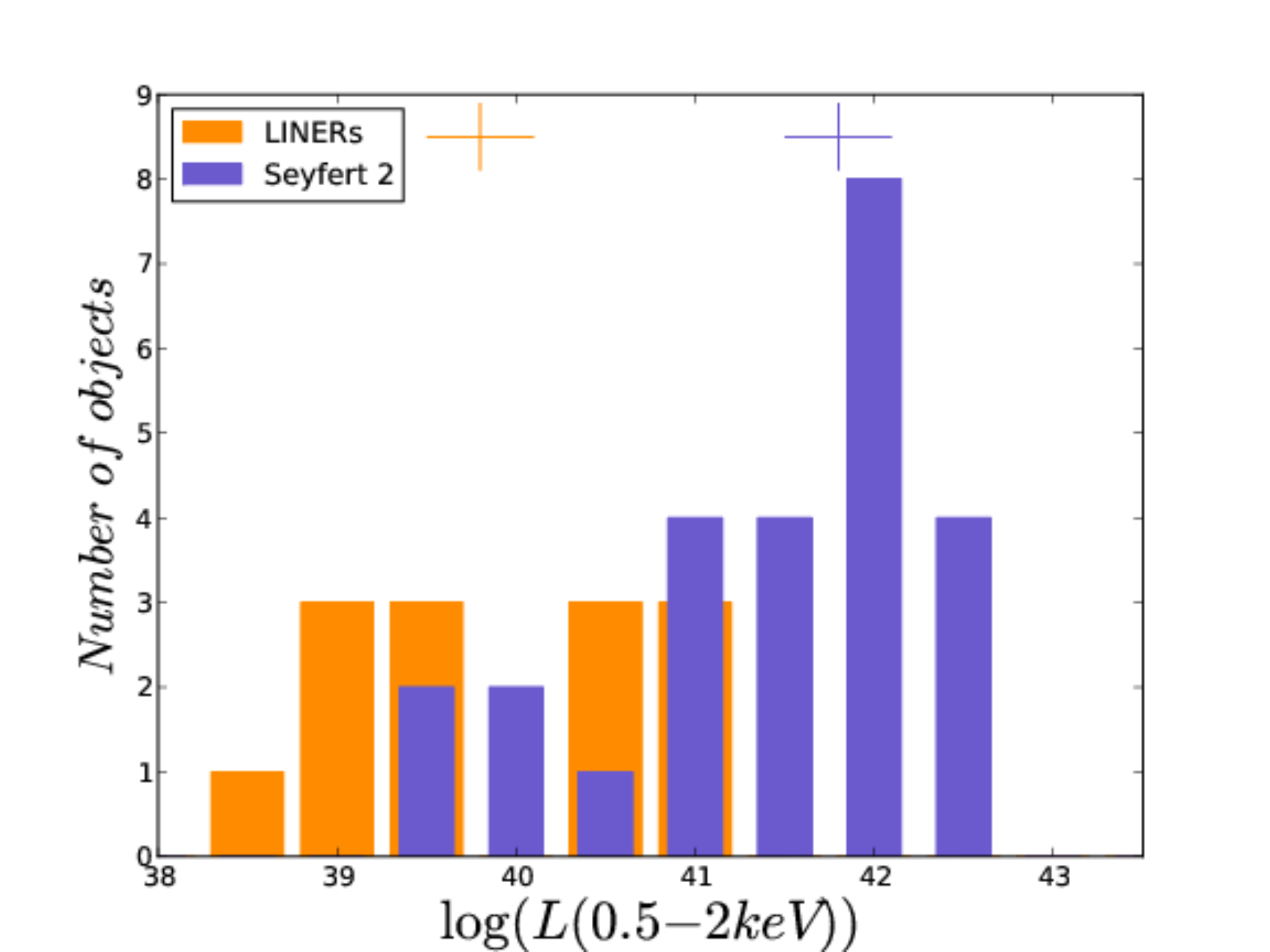}}
{\includegraphics[width=0.4\textwidth]{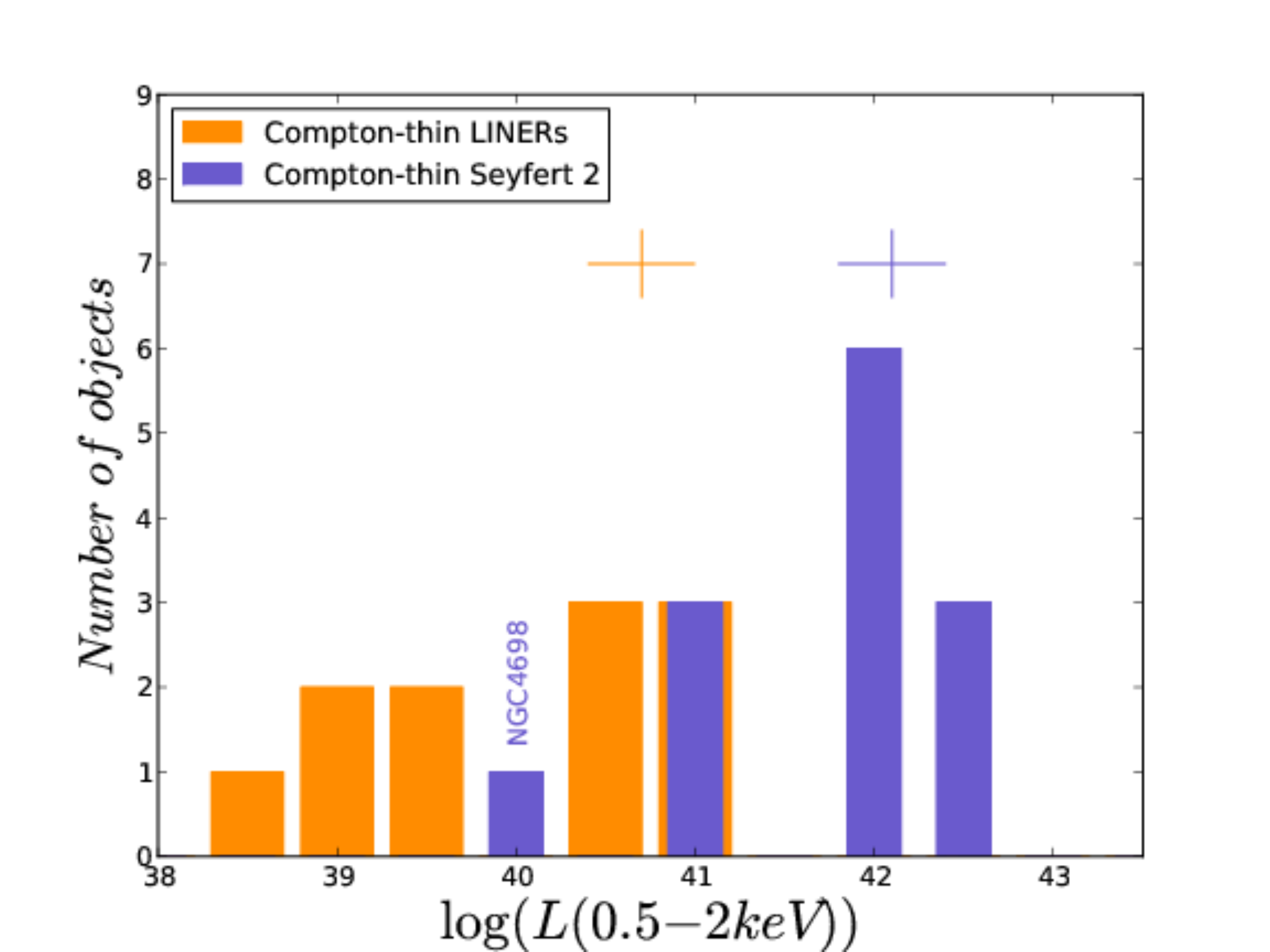}}

{\includegraphics[width=0.4\textwidth]{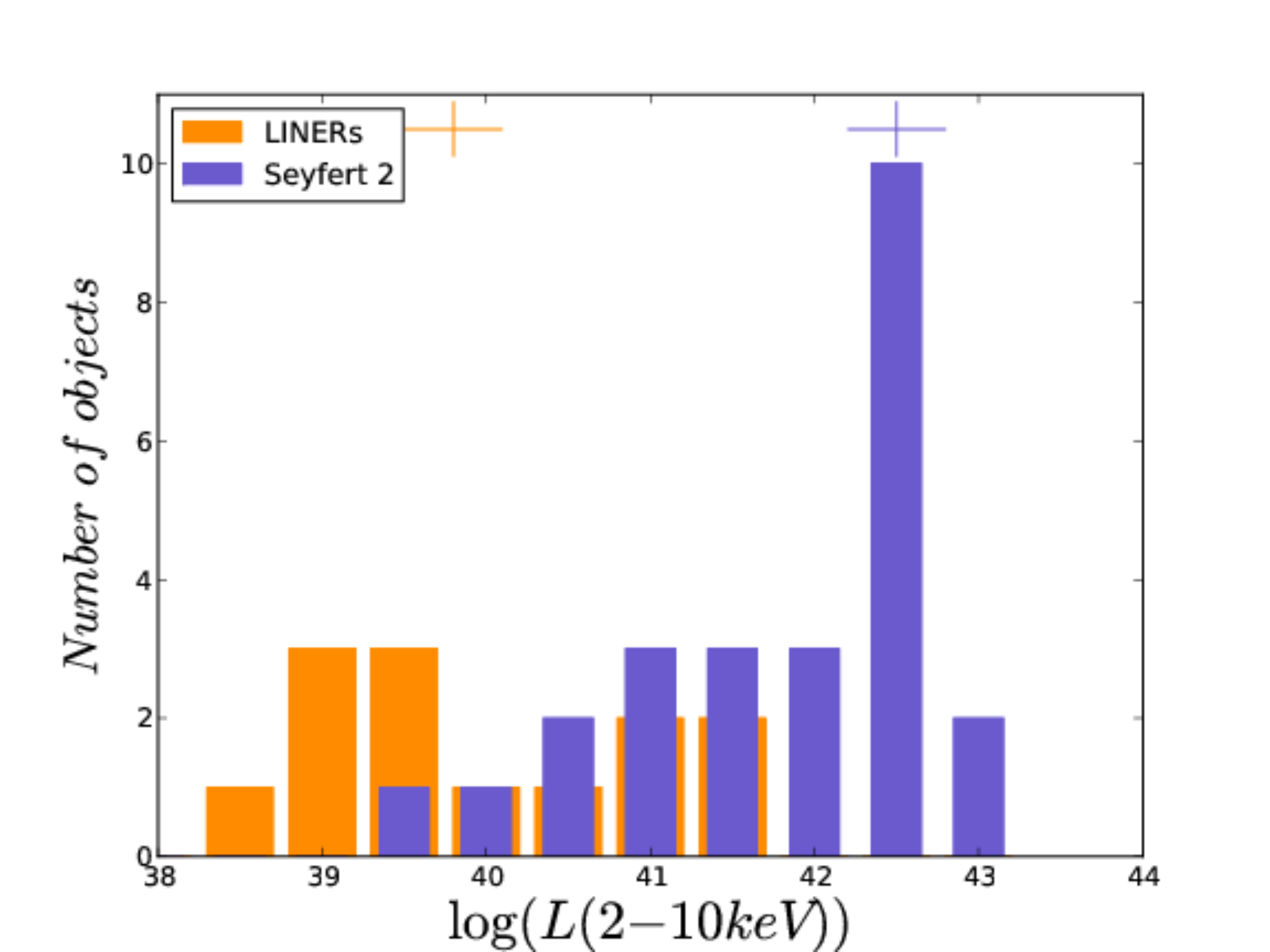}}
{\includegraphics[width=0.4\textwidth]{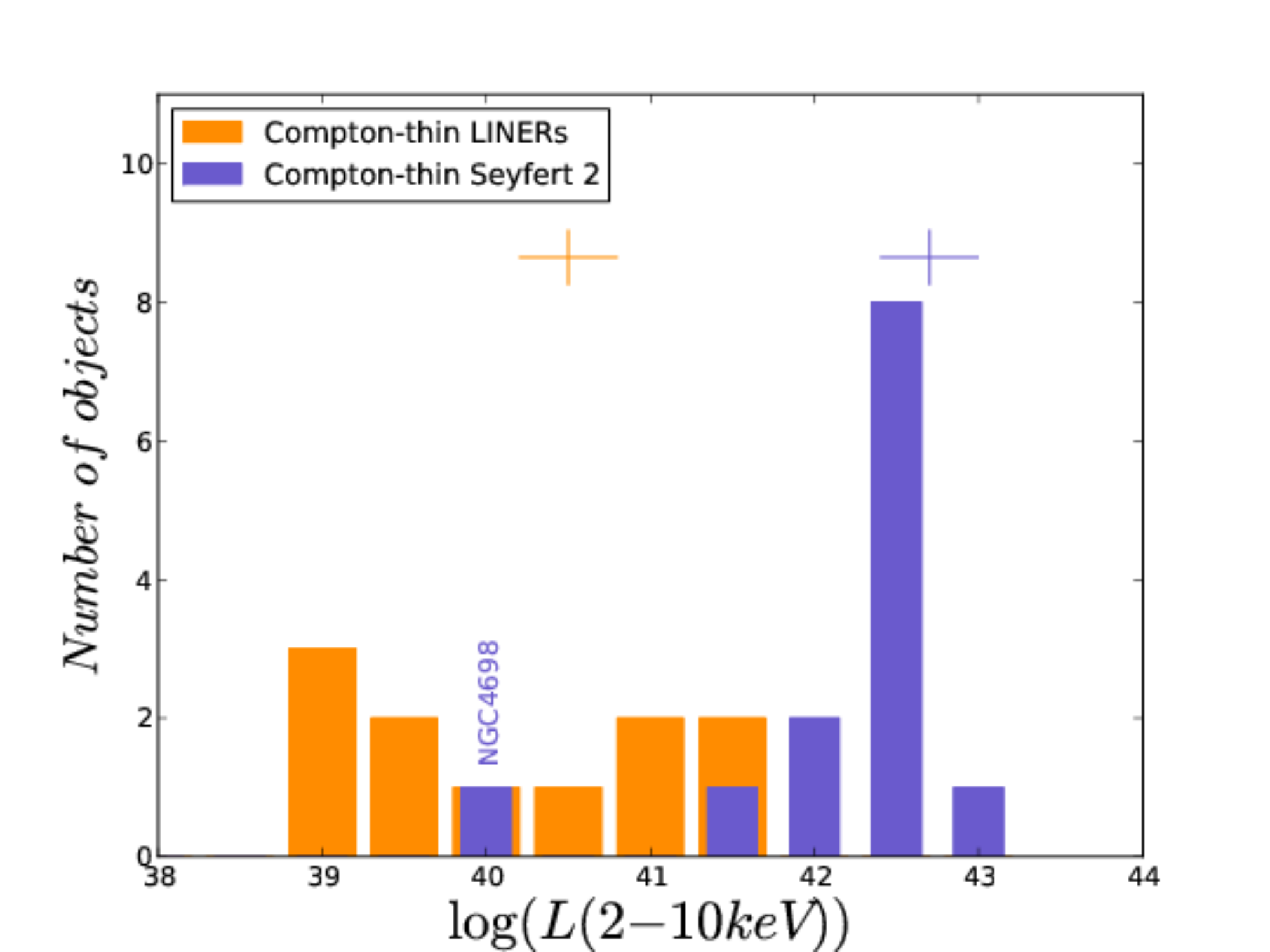}}

{\includegraphics[width=0.4\textwidth]{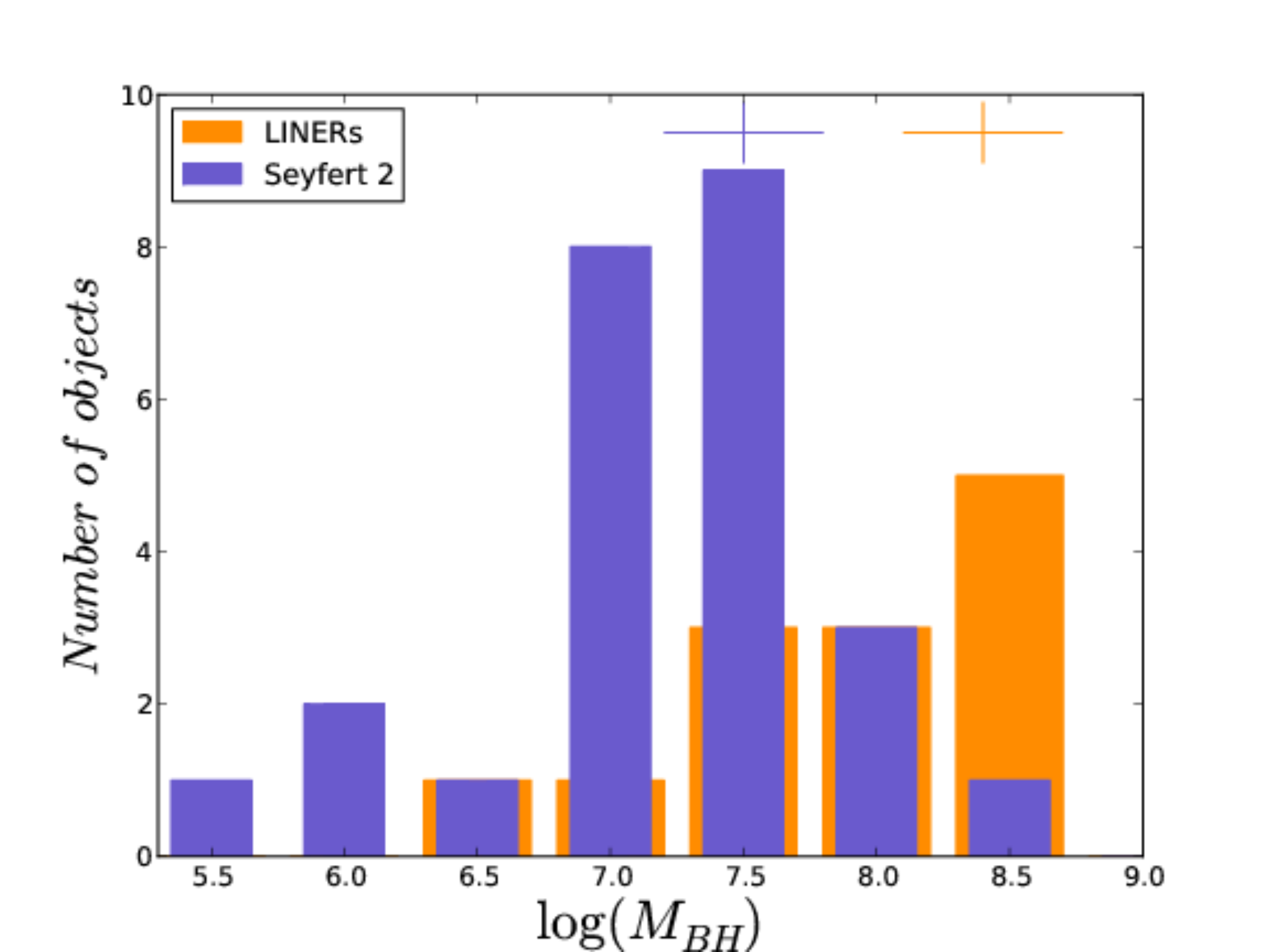}}
{\includegraphics[width=0.4\textwidth]{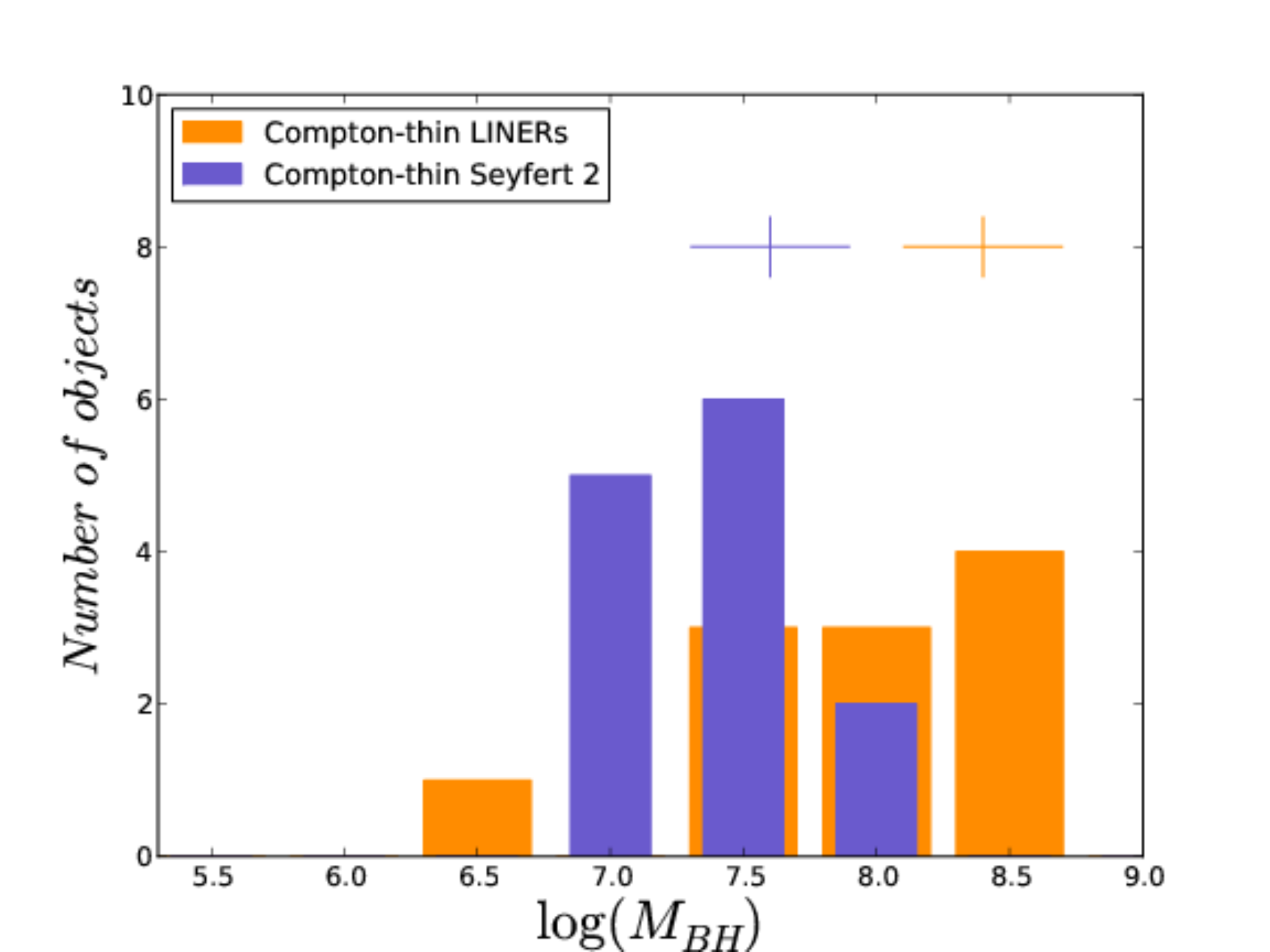}}

{\includegraphics[width=0.4\textwidth]{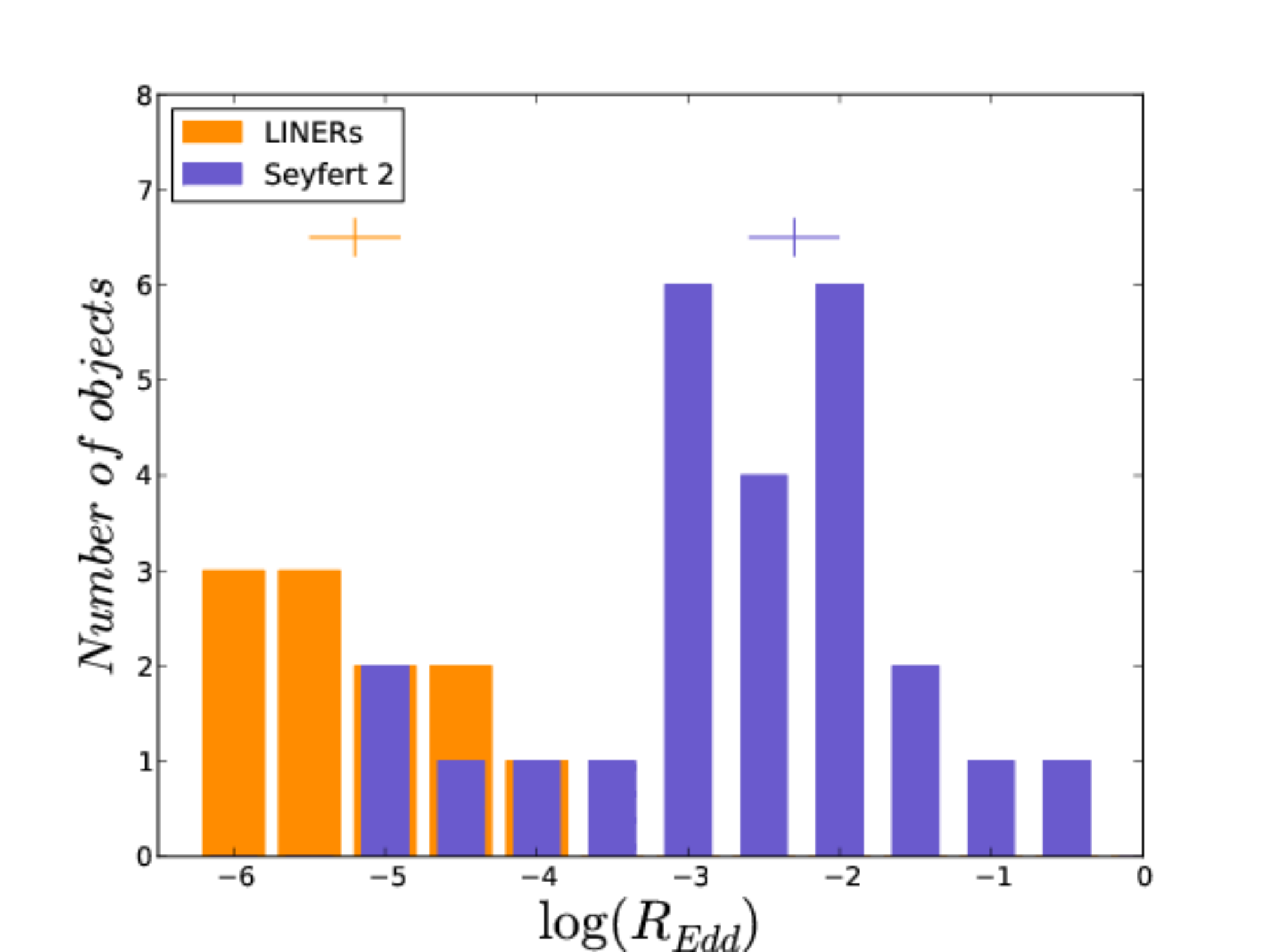}}
{\includegraphics[width=0.4\textwidth]{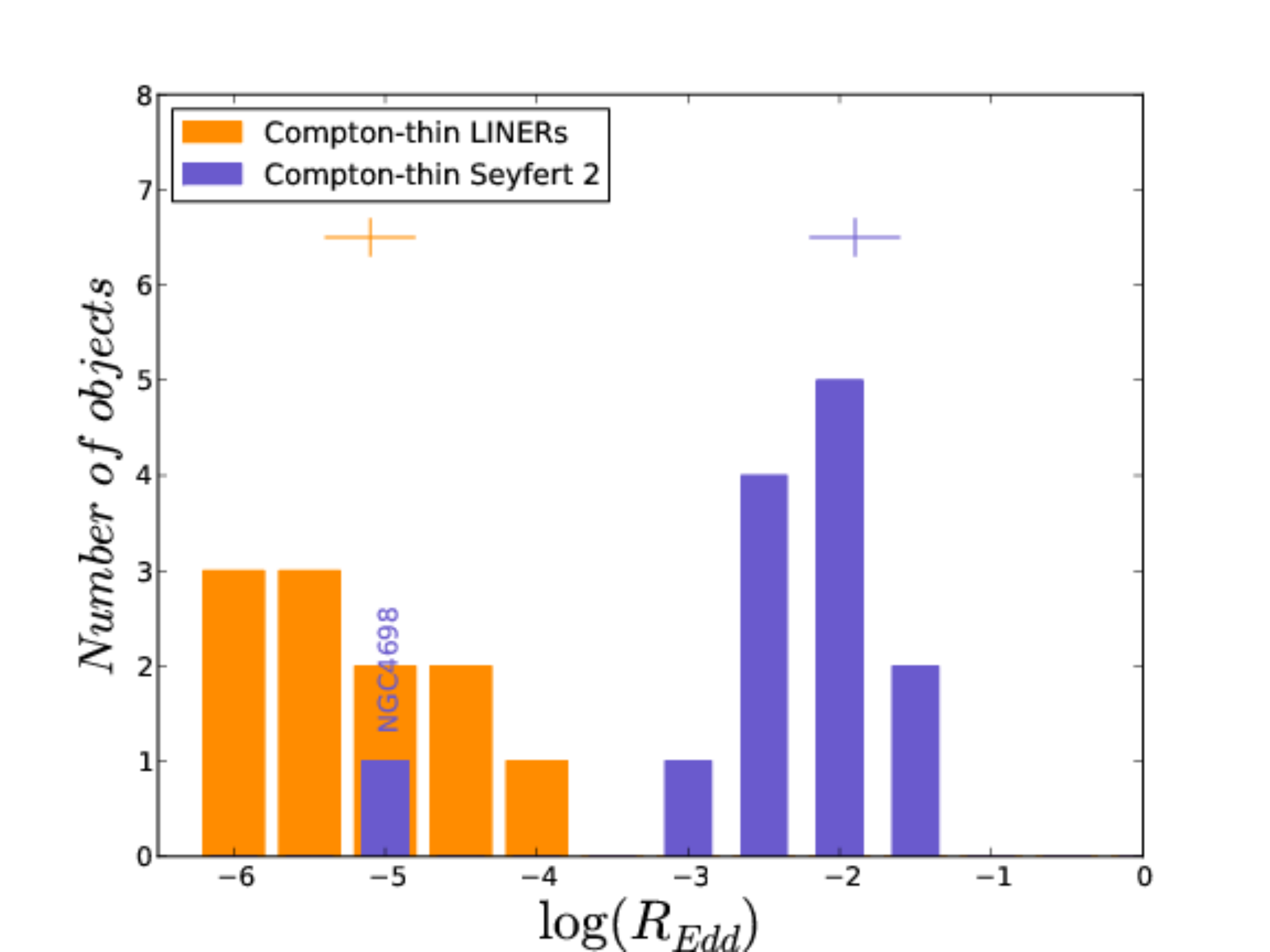}}

\caption{\label{spectralmas}Histograms of the properties of the nuclei. From upper to lower panels: the luminosities in the soft X-ray energy band, $L(0.5--2 keV)$, the luminosities in the hard X-ray energy band, $L(2--10 keV)$, the black hole masses in logarithmic scale, $M_{BH}$, and the Eddington ratios in logarithmic scale, $R_{Edd}$. In all cases (left): all the sample of LINERs and Seyfert 2s, and (right): Compton-thin LINERs and Seyfert 2s. The crosses represent the median values reported in Table \ref{means}.}
\end{figure*}

\begin{figure*}
\centering
{\includegraphics[width=0.49\textwidth]{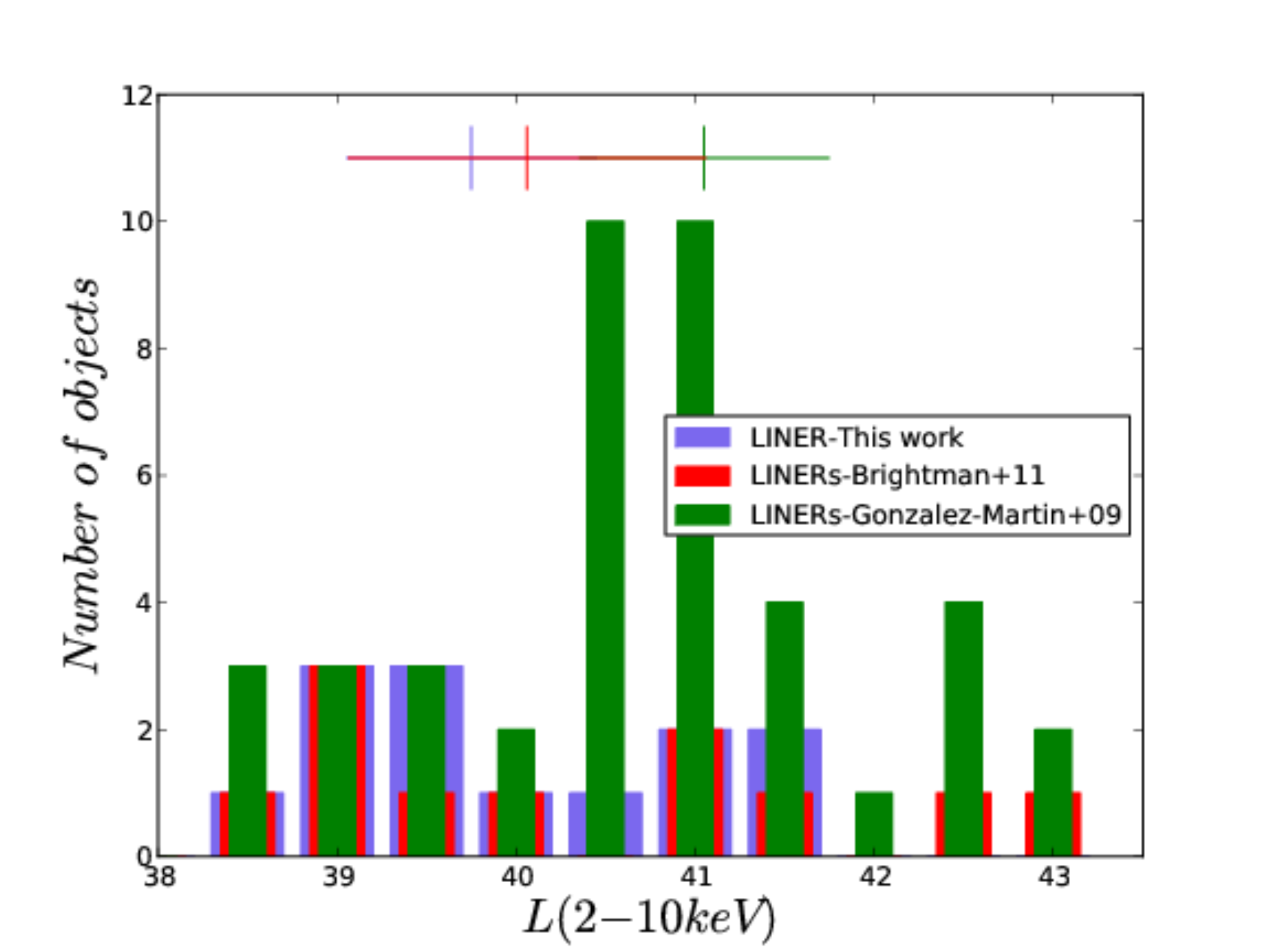}}
{\includegraphics[width=0.49\textwidth]{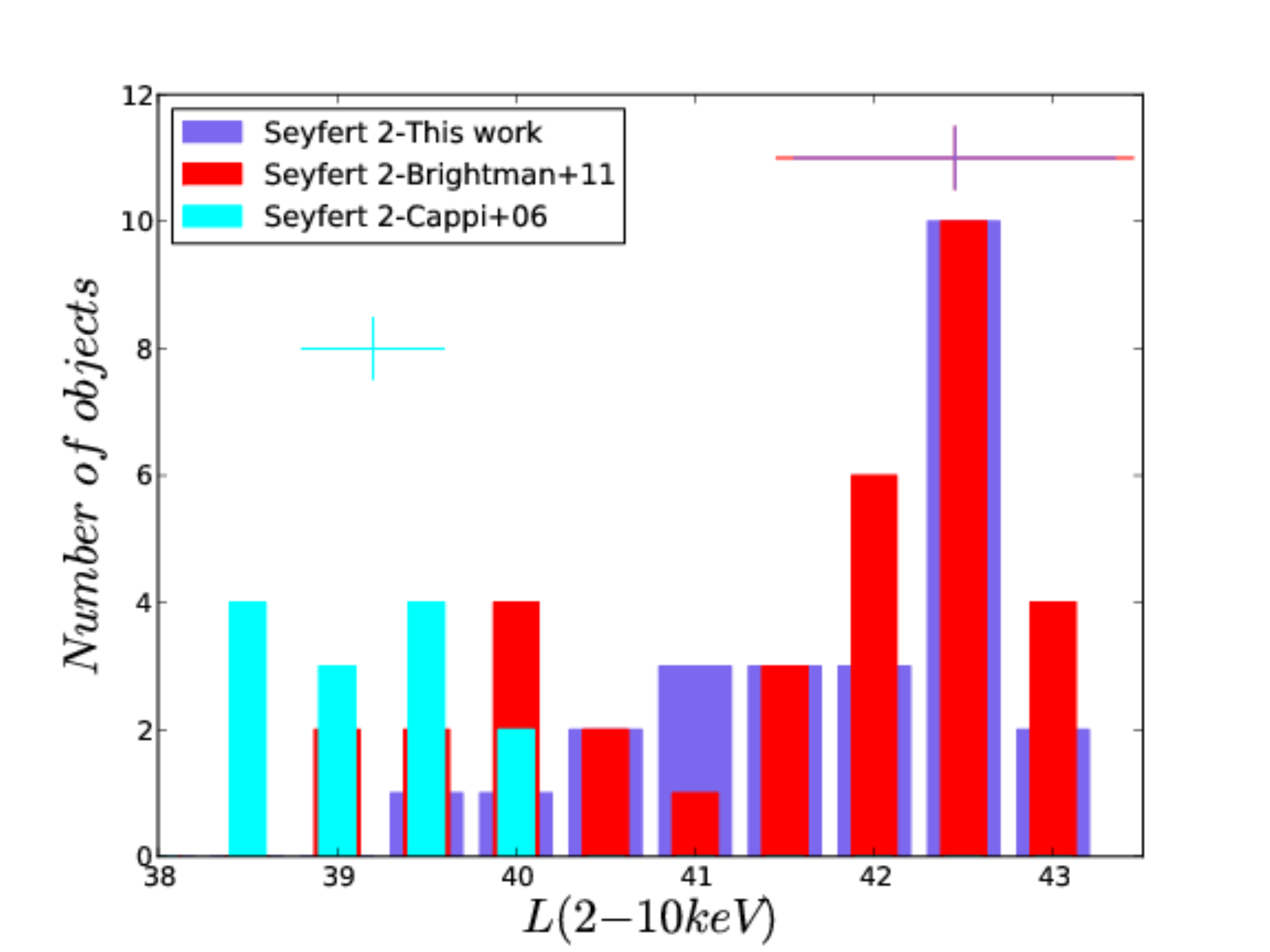}}
\caption{\label{comparisonlxtodo}Histograms of the X-ray luminosities in the 2--10 keV energy band for all the LINERs (left) and Seyfert 2s (right). The histograms include data from \cite{brightman2011} in red, from \cite{omaira2009a} in green, from \cite{cappi2006} and from this work in purple. The crosses represent the median values.}
\end{figure*}

\begin{figure*}
\centering
{\includegraphics[width=0.49\textwidth]{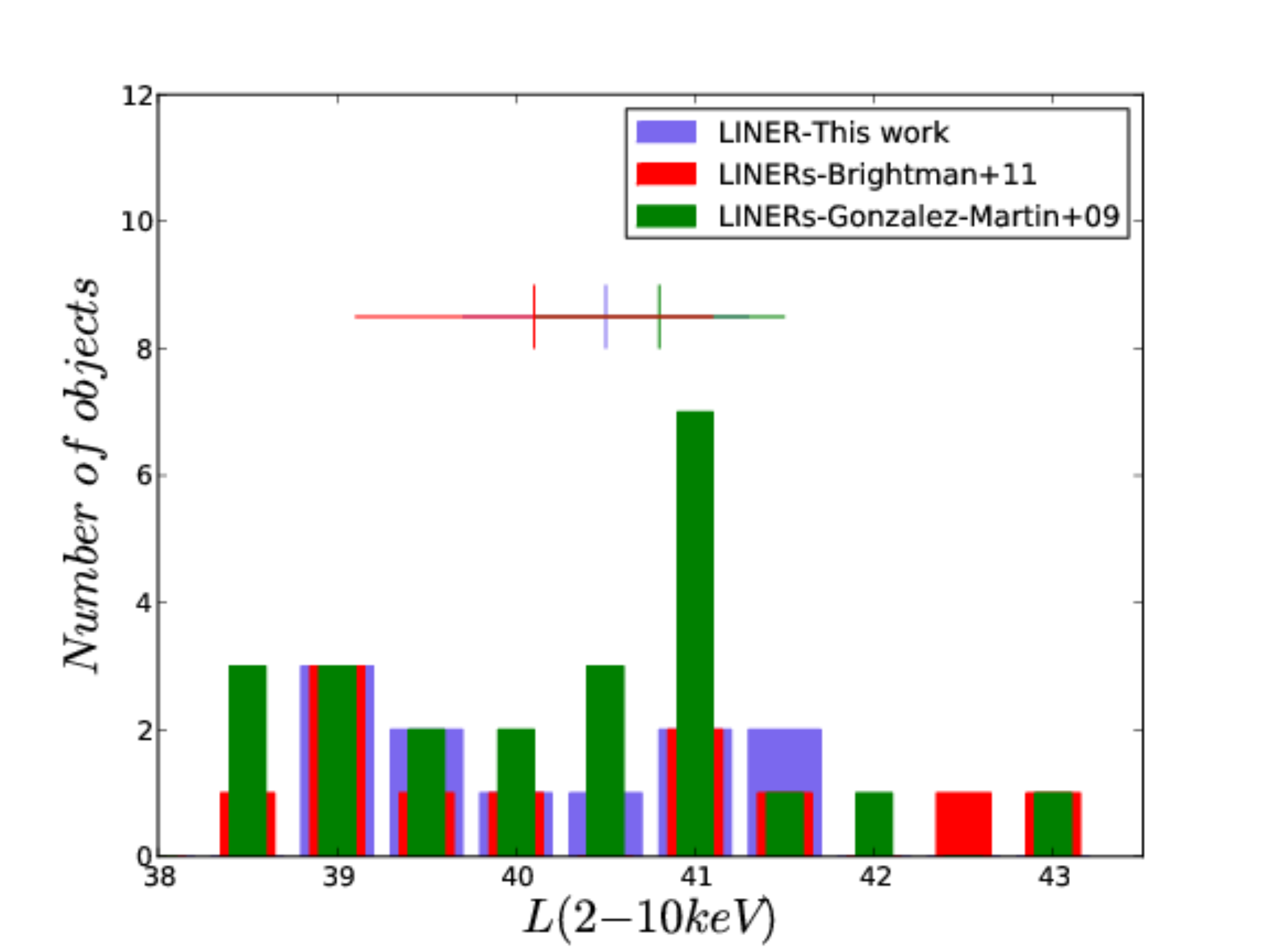}}
{\includegraphics[width=0.49\textwidth]{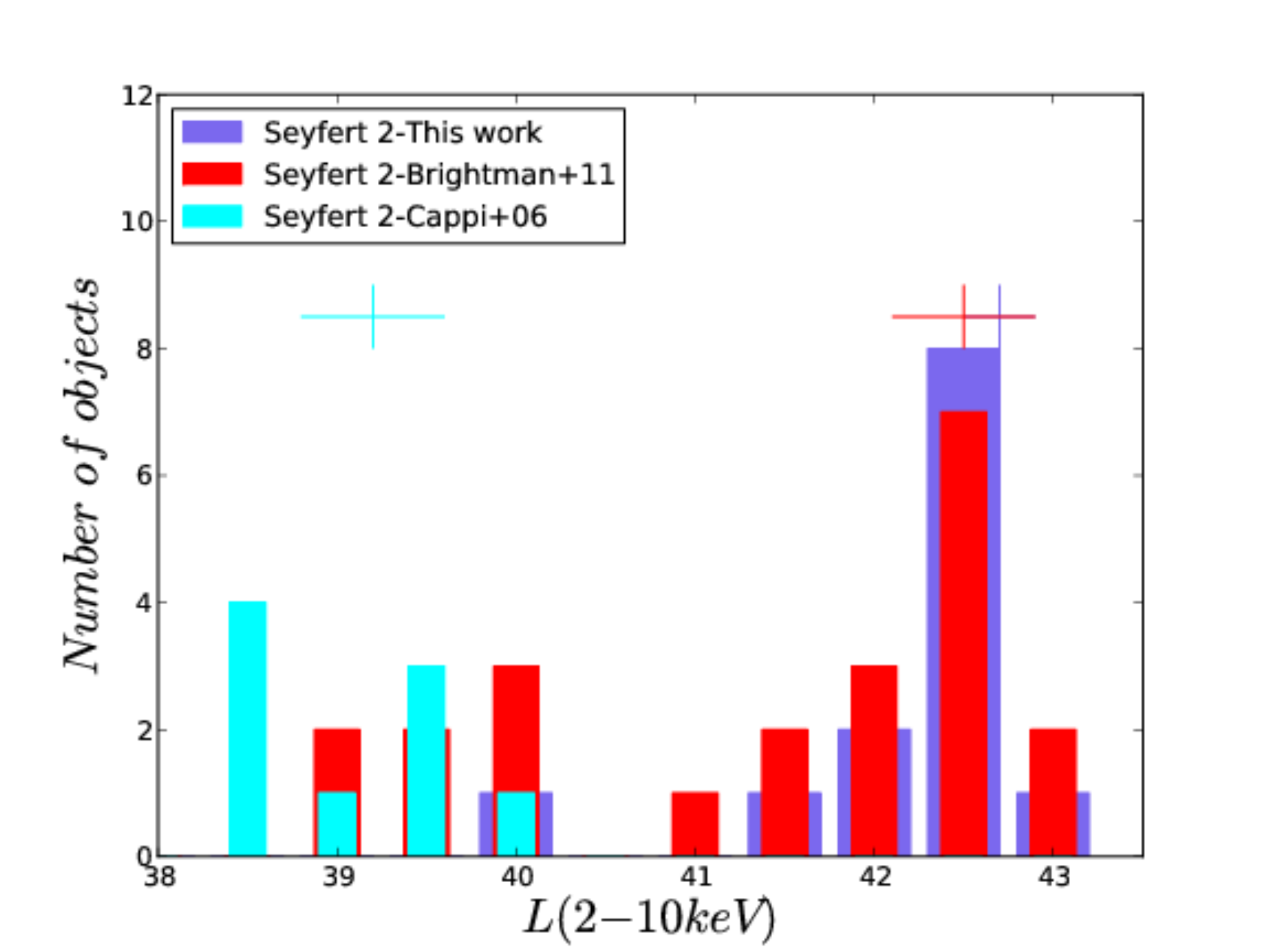}}
\caption{\label{comparisonlx}Histograms of the X-ray luminosities in the 2--10 keV energy band for Compton-thin LINERs (left) and Seyfert 2s (right). The histograms include data from \cite{brightman2011} in red, from \cite{omaira2009a} in green, from \cite{cappi2006} and from this work in purple. The crosses represent the median values.}
\end{figure*}

\begin{figure*}
\centering
{\includegraphics[width=0.49\textwidth]{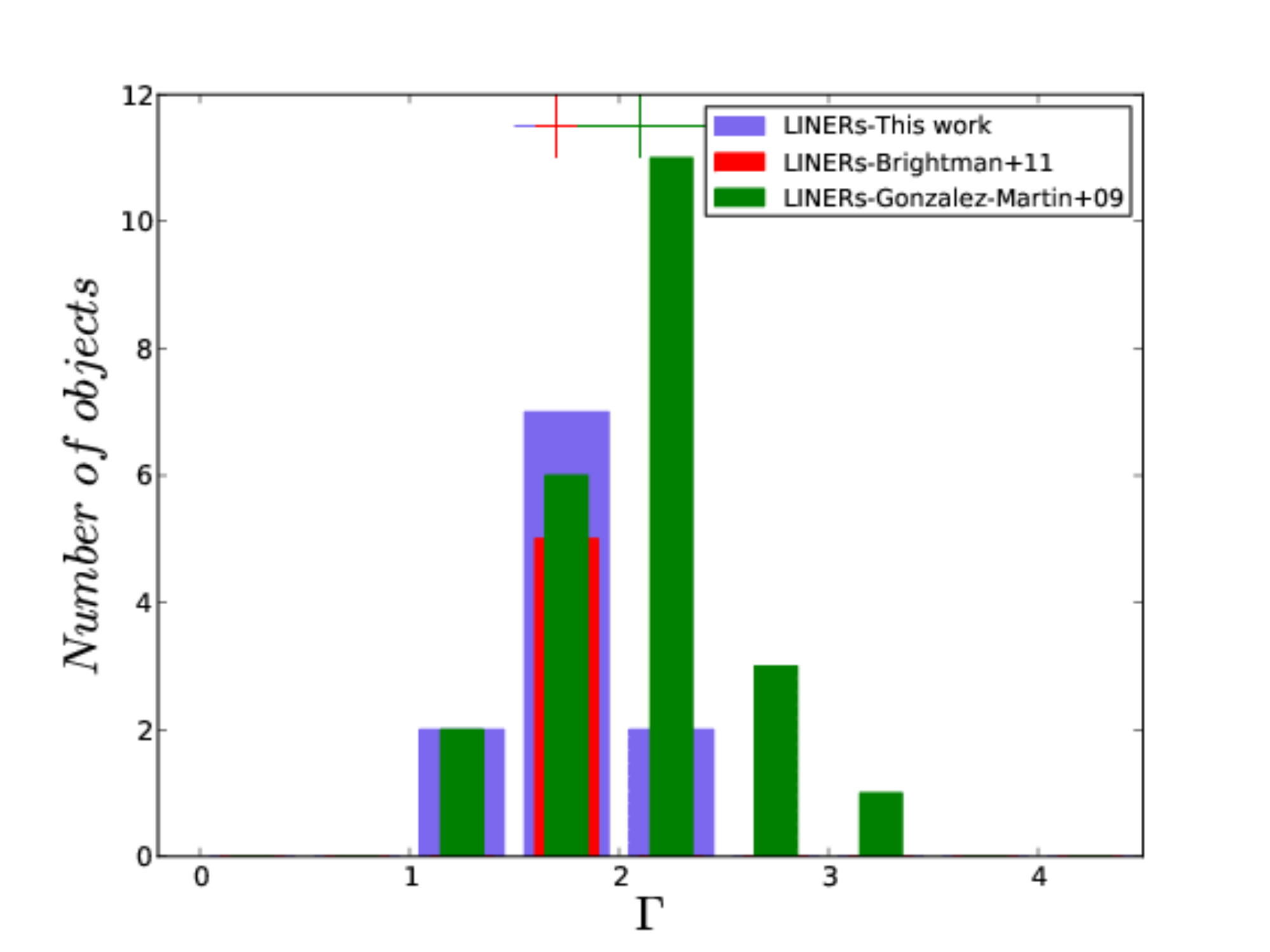}}
{\includegraphics[width=0.49\textwidth]{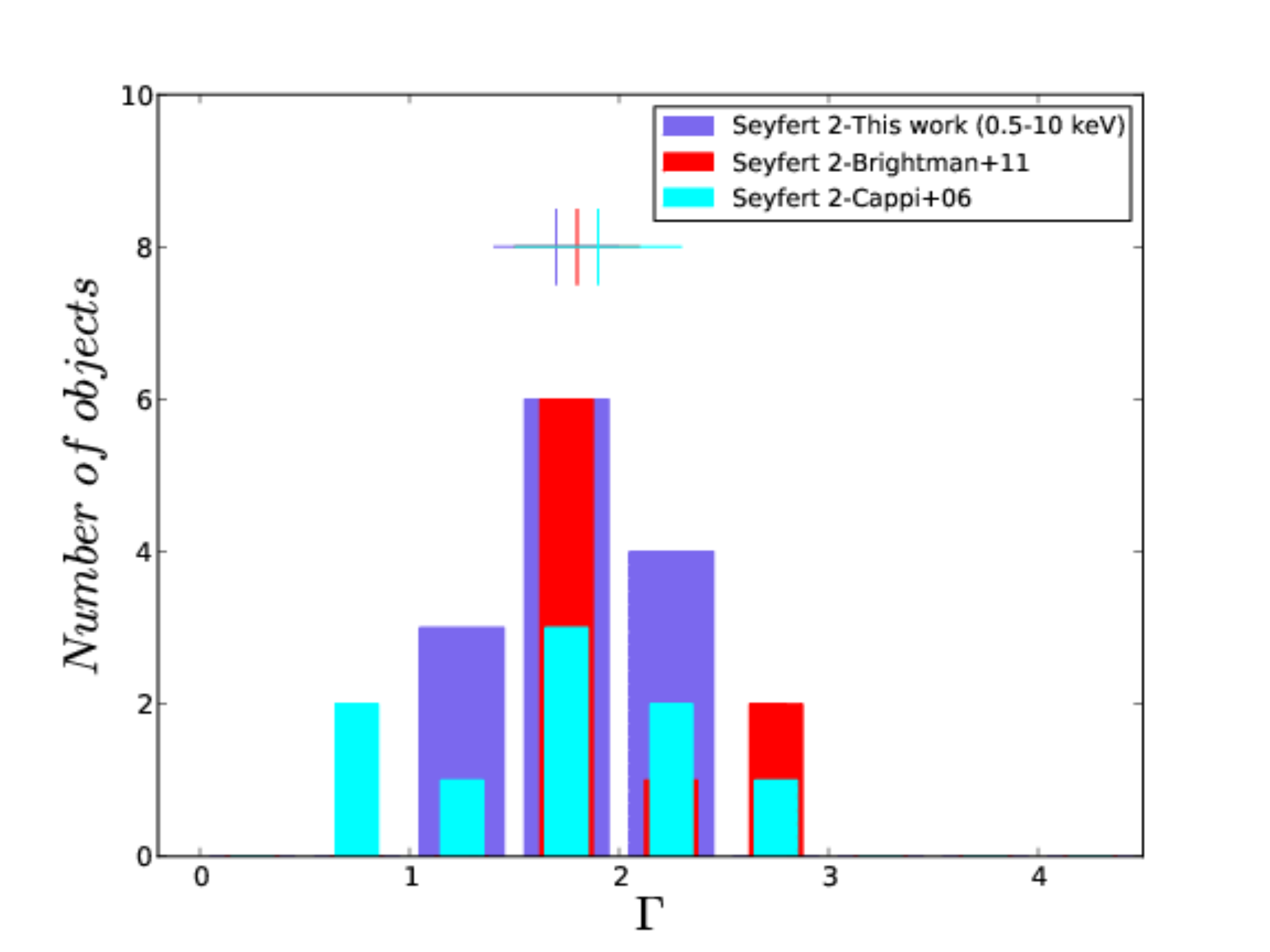}}
\caption{\label{comparisongamma}Histograms of the spectral index for LINERs (left) and Seyfert 2s (right). The histograms include data from \cite{brightman2011} in red, from \cite{omaira2009a} in green, from \cite{cappi2006} and from this work in purple. The crosses represent the median values. }
\end{figure*}


\section{\label{comparison}Results}

\subsection{\label{spectra}Spectral shape and X-ray parameters}

We have compared a sample of 13 LINERs and 25 Seyfert 2s. Among them, two LINERs and 12 Seyfert 2s have been classified as Compton-thick candidates \citep[][HG+15]{omaira2009a}. Observations have shown that the X-ray spectra of these objects are most probably dominated by a reflection component \citep{awaki1991}. Therefore it might happen that the AGN intrinsic nuclear continuum is not accessible at the energies analyzed in this work (0.5--10 keV), what also explains the lack of X-ray variations. For this reason, we have differentiated between Compton-thick and Compton-thin candidates for this analysis.
Here we do not consider the Compton-thick LINERs (see Sect. 2).

Figs. \ref{models} and \ref{spectral} show the main spectral parameters obtained from our analysis, whose median values and the 25\% and 75\% quartiles are presented in Table \ref{means}. In all histograms, the values for the whole
sample are presented in the left panels, and excluding Compton-thick candidates (i.e., only Compton-thin sources) in the right panels, with the median values represented with crosses. 
In order to test whether there are differences between the spectral parameters in LINER and Seyfert 2, we have run a Mann--Whitney U test (U). This is a non parametric test that allows to identify differences between two independent samples, and is appropriate for small samples like ours. When the resulting p-value is smaller than 0.05, we considered that the samples arise from different distributions.

Fig. \ref{models} shows that Seyfert 2s require more complex models to fit their spectra. Specifically, no LINER requires the use of the 2ME2PL model, a difference that is more conspicious when Compton-thick candidates are included (Fig. \ref{models}, left).
From Fig. \ref{spectral} it can be seen that there are no differences between the absorber at soft energies ($N_{H1}$, U test p=0.3),
which is compatible with the Galactic value in most cases, but 
Seyfert 2s appear more absorbed at hard energies ($N_{H2}$, U test p=0.0003). 
It is worth noticing that $N_{H2}$ cannot be well constrained for Compton-thick candidates on the analyzed energies because the spectra are reflection dominated, and the result is that we obtain lower values of $N_{H2}$ of the order of $10^{23} cm^{-2}$ instead of the required $>10^{24} cm^{-2}$ in Compton-thick sources \citep[see e.g.][]{brightman2011}.
The indices of the power law representing the AGN are very similar in both families, although Compton-thick candidates show much flatter values, as expected \citep[e.g.,][]{cappi2006}. When only Compton-thin are considered, both distributions look pretty similar (p=0.4). Finally, a clear difference between the objects is observed in the temperatures of the thermal component, with Seyfert 2s showing a bimodal distribution with medians at $kT$ = 0.1 keV and $kT$ = 0.7 keV, whereas LINERs show a peak on the distribution centre at $kT$ $\sim$ 0.6 keV. For this reason, both distributions cannot be directly compared through the U test. For the $kT$ distribution of Seyfert 2s, we have run a Kolmogorov-Smirnov (K-S) test to check whether it comes from a normal distribution; this results in a p-value of 0.0003, rejecting the null hypothesis. Therefore our results show that Seyfert 2s show a bimodal distribution in $kT$.

In Fig. \ref{spectralmas}, X-ray luminosities from the fitted models, black hole masses, $M_{BH}$, and Eddington ratios, $R_{Edd}$, are presented. $M_{BH}$ have been calculated from the $M_{BH}$--$\sigma$ relation \citep[][$\sigma$ from HyperLeda]{tremaine2002} or taken from the literature otherwise (see HG+14; HG+15). $R_{Edd}$ are obtained differently from our previous works, using a bolometric correction, $k$, dependent on luminosity instead of a constant value. We calculated $R_{Edd}$ following \citet{eracleous2010} and $k$ following \cite{marconi2004}. From their $k$--$L_{bol}$ relation, we determined the
$k$--$L(2-10$ keV$)$, and fitted a fourth order polynomial to derive the relation, which was applied to each object. 
Seyfert 2s show lower values of $M_{BH}$ than LINERs, although with a substantial overlap (U test, p=0.008). Seyfert 2s show an order of magnitude higher soft (0.5-2.0 keV, U test p=0.0001) and hard (2-10 keV, U test p=$7 \times 10^ {-5}$) X-ray luminosities.
A clear difference is found for $R_{Edd}$ at $\sim 10^{-3}$, with Seyfert 2s (LINERs) located above
(below) this value (U test p=$6 \times 10^ {-5}$). Note that only one Seyfert 2, namely NGC\,4698, appears in luminosity and $R_{Edd}$ with a typical value of what it is found for LINERs. The optical
classification of this object has been controversial in the current literature, being classified as a Seyfert 2 by \cite{ho1997} and \cite{bianchi2012}, but also classified as a LINER by \cite{omaira2009a}. However, since it is not a variable source, its presence in any of the two families does not change our conclusions concerning the variability patterns.

Thus, we find that major differences are reported in the X-ray luminosities and the Eddington ratios. To test the reliability of this result we performed a clustering analysis (see Appendix), which reveals that we can divide the sample in four groups (LINER A, LINER B, Seyfert 2 A, and Seyfert 2 B). These can be differentiated based on three spectral parameters; the spectral index, the X-ray luminosity and the Eddington ratio, in very good agreement with our results.

\subsection{\label{previous}Comparison with previous results} 

We have compared our values of the intrinsic properties, $\Gamma$ and L($2-10$ keV), of LINERs and Seyfert 2s with previous works. The most complete sample in the literature is the 12 $\mu$m galaxy sample presented by \cite{brightman2011}, which includes 37 Seyfert 2s and 11 LINERs. Among these, they classified 13 Seyfert 2s as Compton-thick sources, whereas none of the LINERs showed Compton-thickness indications. This classification was based on the characterization of high $N_H$, that they were able to estimate using a model of a spherical distribution of matter, and because the source spectrum was dominated by a large reflection fraction. 
We have also included the sample by \cite{cappi2006}, which includes the 30 Seyferts from the Palomar sample \citep{ho1997} located at distances below 22 Mpc. Among these, 13 are Seyfert 2s; four of them classified as Compton-thick candidates using the same indicators we used.
We have also added the sample by \cite{omaira2009a} that contains 82 LINERs. This is the largest sample of LINERs that has been studied at X-rays in the literature. From their sample, we removed those sources classified as non-AGN candidates, as we did with our sample (see Sect. 2). 
We compared the spectral fits of the common sources from all these works and found that the spectral parameters and luminosities are consistent between the different analyses.

Figs. \ref{comparisonlxtodo} (all the sources) and \ref{comparisonlx} (Compton-thin sources only) show the histograms of the hard X-ray luminosity for LINERs (left) and Seyfert 2s (right). The median values [25\% and 75\% quartiles] for the Compton-thin LINERs are log(L(2--10 keV)) = 40.1[39.2--41.7], 40.8[39.9--41.3], and 40.5[39.5--41.2] [erg/s] for the samples by \cite{brightman2011}, \cite{omaira2009a}, and this work. It can be seen that most of the LINERs are located log(L(2--10 keV))$<$42 [erg/s], with only four sources above this value. The median values [25\% and 75\% quartiles] for the Compton-thin Seyfert 2s are log(L(2--10 keV)) = 39.2[38.9--39.6], 42.5[40.2--42.9] and 42.7[42.5--42.8] [erg/s] for the samples by \cite{cappi2006}, \cite{brightman2011} and this work. The work by \cite{cappi2006} shows $\sim$ three orders of magnitude lower luminosities. This difference can be explained by selection effects. Whereas their sample include sources located below 22 Mpc, ours include only four Seyfert 2s within this distance range. Therefore, it could be possible that our selection criteria of having at least two observations to be used for the analysis results in the selection of brighter Seyfert 2s. As it can be seen in Fig. \ref{comparisonlx}, the sample by \cite{brightman2011} includes sources in the whole luminosity range. Therefore, we cannot conclude that LINERs have lower X-ray luminosities than Seyfert 2s, since they show an overlap in the  log(L(2--10 keV))=38--42 [erg/s] luminosity range. Conversely, there are no LINERs showing luminosities higher than log(L(2--10 keV))=42 [erg/s], which is confirmed by the largest sample \citep{omaira2009a}.

Fig. \ref{comparisongamma} shows the histograms of $\Gamma$ only for the Compton-thin\footnote{Whereas our values of $\Gamma$ for Compton-thick sources were obtained from a spectral fit in the 2--10 keV energy band, \cite{cappi2006} calculated them in the 0.5-10 keV energy band, thus the values cannot be directly compared.} LINERs (left) and Seyfert 2s (right). \cite{brightman2011} fixed the value of $\Gamma$ = 1.9 when it was not well constrained, so we did not take those sources into account. The median values [25\% and 75\% quartiles] for the LINERs are $\Gamma$ = 1.7[1.6--1.8], 2.1[1.7--2.4], and 1.7[1.5--1.9] for the samples by \cite{brightman2011}, \cite{omaira2009a}, and this work. The median value obtained by \cite{omaira2009a} is steeper but compatible within the errors. The median values [25\% and 75\% quartiles] for the Seyfert 2s are $\Gamma$ = 1.9[1.3--2.0], 1.8[1.7--2.1] and 1.7[1.5--2.0] for the samples by \cite{cappi2006}, \cite{brightman2011} and this work. 
We notice that two sources in the sample of \cite{cappi2006} have $\Gamma < 1$ (NGC\,2685 and NGC\,3486), which have been classified as Compton-thick candidates in the literature \citep{omaira2009b,annuar2014}.
Therefore, our results are in good agreement with the spectral parameters reported before. 
These results show that $\Gamma$ in LINERs and Seyfert 2s are found in the same range, but the confidence intervals are so large that different astrophysical scenarios could explain them.

Our results are in agreement with previous works. We note however that our selection criteria of having more than one observation per source leaves the faintest Seyfert 2s out of our sample.

\begin{figure}
\centering
{\includegraphics[width=0.49\textwidth]{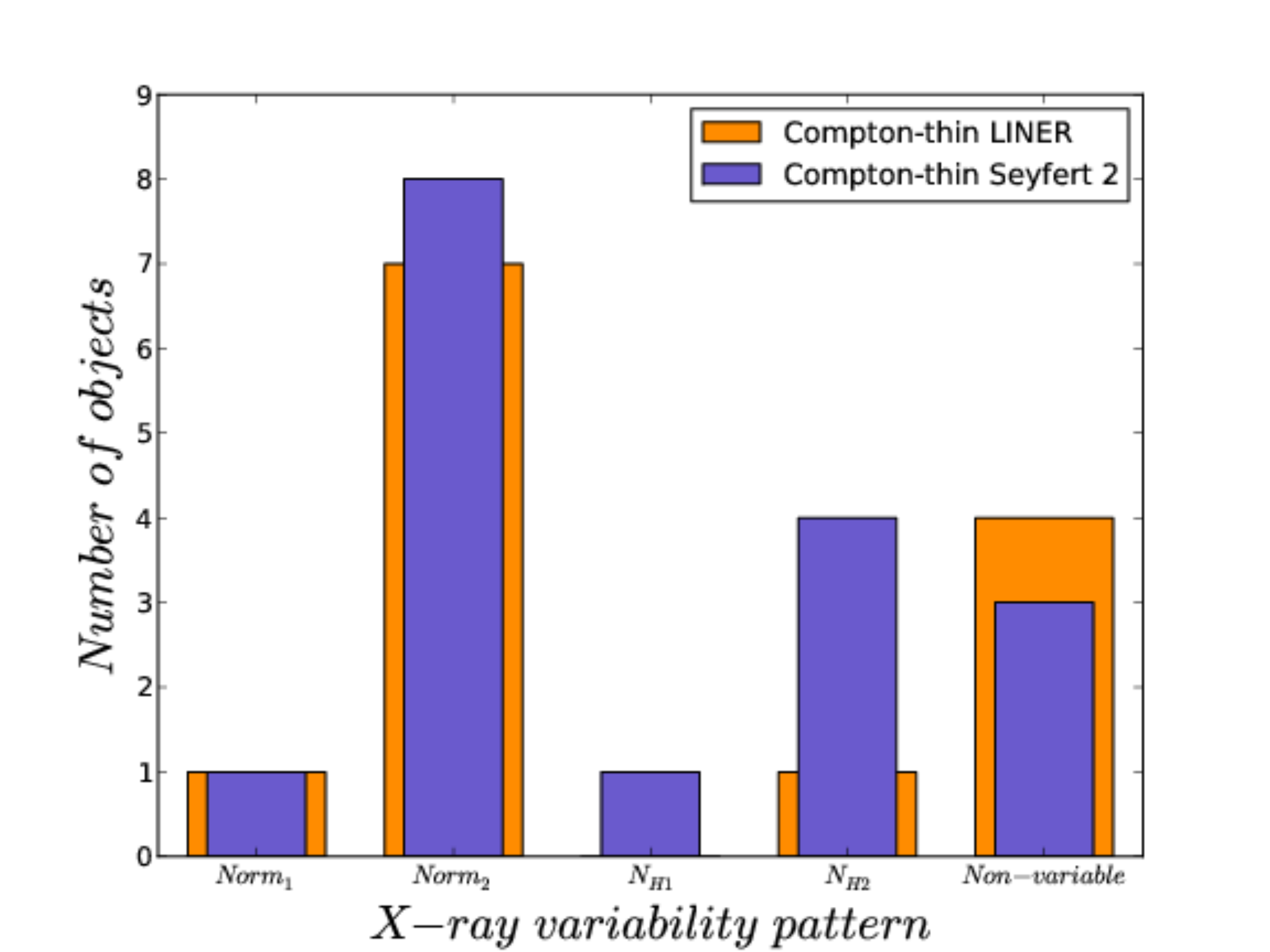}}
\caption{\label{patterns}Histograms of the X-ray variability patterns of the Compton-thin LINERs, and Compton-thin Seyfert 2s. These are the variable parameters that refer to the normalizations at soft ($Norm_1$) and hard ($Norm_2$) energies, and/or the absorber at soft ($N_{H1}$) and hard energies ($N_{H2}$). The $non-variable$ cases refer to those nuclei where variations were not detected. Note that in a few sources two parameters vary together (see Table \ref{properties}).}
\end{figure}

\subsection{X-ray and UV variability}

X-ray short-term variations cannot be claimed in any of the studied
objects, as all the measurements were below the 3$\sigma$ level.
Regarding long-term variations, it is found that LINERs and Seyfert 2s are X-ray variable objects in timescales ranging from months to years, except when they are Compton-thick objects, where variations are not usual (HG+15).
When a source was observed in a Compton-thick state in one spectrum and in a Compton-thin state in another spectrum, we classified it as a changing-look candidate; four changing look candidates are included in the sample of Seyfert 2s, from our own analysis (MARK\,273 and NGC\,7319; HG+15) or taken from the literature (MARK\,1210, \citealt{guainazzi2002}; and NGC\,6300, \citealt{guainazzi2002a}).
Using the same indicators as in HG+15\footnote{$\Gamma<1$, $EW(Fek\alpha)>500eV$, and $F(2-10 keV)/F([OIII])<1$.}, we did not find changing-look candidates among LINERs. 

The histogram of the X-ray variability patterns is presented in Fig. \ref{patterns}. The most frequent long-term variations observed in both families of AGN are related to the normalization of the power law at hard energies, $Norm_2$, which are observed in all the eight variable LINERs and in nine out of the 11 variable Seyfert 2s\footnote{Note that one LINER and one Seyfert 2 are Compton-thick candidates and thus are not counted in the histogram.} with amplitudes ranging from 20\% to 80\%. Variations due to absorption are less common, being more frequent in Seyfert 2s (four out of 11, i.e., 36\%) than in LINERs (one out of eight, i.e., 13\%). Variations at soft energies (i.e., in $Norm_1$ or $N_{H1}$) are found in only two Seyfert 2s and one LINER, in all cases accompanied with variations of the nuclear continuum. Variations at soft X-ray energies are rare and should be confirmed with more data before its discussion (HG+15).

The last result we like to report is that at UV frequencies long-term UV variations are common in LINERs (five out of six), whereas this kind of variations are not detected in Seyfert 2s. It is worth noting that the presence of the nuclear UV source in Seyfert 2s is very scarce, as we have detected it only in three cases, among the nine sources where UV data were available.


\section{\label{discusion}Discussion}

LINERs have been invoked as a scaled down
version of Seyfert galaxies based in their average luminosities and $R_{Edd}$ \citep{omaira2006, omaira2009b, omaira2009a, younes2011}. \cite{omaira2009a} pointed to the overlap found in their properties taking the Seyfert sample from \cite{panessa2007} as a reference. A drawback of all these works could be that they also include Compton-thick objects. 
In HG+15 we found that Compton-thick sources do not vary at X-rays, in agreement with other works \citep[e.g.,][]{lamassa2011,arevalo2014b}. Since Compton-thick sources seem to be dominated by a reflection component \citep[e.g.,][]{awaki1991}, our results show that this component remains constant with time, most probably because it is located far from the SMBH.
We know that the 2--10 keV energy band in Compton-thin and Compton-thick nuclei is dominated by a direct and a reflection component, respectively. Thus, the present work separating Compton-thin and Compton-thick sources allows a more net view on the nature of these families. In the following we analyze the accretion state and the nature of the torus and the broad line region (BLR) of LINERs and Seyfert 2s.

\subsection{Accretion}

It is well constrained that Galactic X-ray binaries (XRBs) show different spectral states and accretion mechanisms \citep{remillard2006}, which present variations in $\Gamma$ and $R_{Edd}$, being a negative correlation between these parameters attributed to inefficient accretion and a positive one to efficient accretion \citep[e.g., Cygnus X-1,][]{ibragimov2005}. 
In analogy with this behaviour, several authors have argued that the accretion mechanism might be different for low and high accreting AGN \citep{lu1999,shemmer2006,ho2008,gucao2009,younes2011,yang2015}. It has been suggested that LINERs are dominated by radiative inefficient accretion flows (RIAF) while Seyferts have a standard accretion disc \citep[see][for a review]{ho2008}.
To date, however, the $\Gamma$--$R_{Edd}$ relation has been reported only in one AGN individually \citep{emmanoulopoulos2012}. Indeed, they used \emph{RXTE} data monitoring the source in timescales of days, and we notice that the variations in $\Gamma$ were of very short amplitude (changes from $\sim$ 1.8 to 1.9). 
Instead of the study of $\Gamma$ variations with $R_{Edd}$ for a single
object, several studies have reported the $\Gamma$--$R_{Edd}$ correlation with
different objects \citep{shemmer2006,gucao2009,younes2011,yang2015}.
We plotted this relation for our LINER and Seyfert 2s (see Fig. \ref{gdibu}), but we cannot confirm any correlation, even if there exists a trend to correlate (anticorrelate) for Compton-thin Seyfert 2s (LINERs).
This result is in agreement with previous works where the $\Gamma$--$R_{Edd}$ produces a very large scatter instead of a clean correlation \citep[e.g.,][HG+13]{gucao2009, yang2015}. 
We find that the main difference between Seyfert 2s and LINERs comes from the luminosity which in turn leads to a higher $R_{Edd}$ for Seyfert 2s (see Fig. \ref{spectralmas}).  
Whereas this is the main difference in the spectral parameters, we do not find variations in $\Gamma$ neither for LINERs nor for Seyferts, as is the case of Galactic X-ray binaries. Thus, either we are not able to recover such a small variations or variations in $\Gamma$ are not the general trend in AGN.

In this work it is reported for the first time that, regardless the LINER or Seyfert nature
of the source, most of the objects show variability in the continuum
normalization, i.e., the transmitted continuum flux from the AGN. Moreover, the
amplitudes and timescales of the variations are similar for both families. 
Since the AGN continuum at X-rays comes from the Comptonization of photons from the inner parts of the accretion disc \citep{shakura1973}, the mechanism driving these variations might be related to
fluctuations in the inner accretion disc. In principle, one might think that this is
against the idea that LINERs are in a different accretion state where the disc
is partially suppressed and RIAFs \citep[][]{quataert2004} take place for the relevant accretion mechanism. However, it is worth noting that these kind of intrinsic
continuum flux variations could be produced by both emission mechanisms, and can be explained in terms of propagating viscous fluctuations in the accretion rate \citep{lyubarskii1997}.
This model is also able to explain X-ray variations in XRBs \citep[e.g. Cyg X-1,][]{arevalouttley2006}.
Thus, we conclude that regardless the accretion mechanism of the nuclei, the observed X-ray variations might be related to propagating viscous fluctuations in the accretion rate.

\subsection{Torus and BLR}

Theoretical works show that at bolometric luminosities below $L_{bol} \sim  10^{42}$
$erg$ $s^{-1}$, the accretion onto the SMBH cannot longer sustain the required cloud
outflow rate, and the torus and the BLR might disappear \citep{elitzur2006}. Our
analysis could bring some light into observational evidences of the lack
of the BLR and/or torus. 

In type 2 sources we do not expect to see the UV continuum of the nuclear source because, from the point of view of the UM, 
it is blocked by the torus. However, when the torus is no longer in our line of
sight (or it is not present at all), we do expect to detect a point like source
associated with the accretion disc. If that is the case, the UV continuum should
show variations because the accretion disc is highly variable \citep[e.g.,][]{done2007}. 
 
Indeed, UV variations are not detected among Seyfert 2s (HG+15), in agreement with
the idea that the torus is in our line of sight for this type of AGN. 
In fact, Seyfert 2s show a resolved nucleus in \emph{HST} data and their UV extended emission is dominated by emission of the ionized gas and/or emission from the underlying galactic bulge \citep{munozmarin2007,munozmarin2009}. On the contrary, \emph{HST} data of LINERs show unresolved point sources at UV frequencies \citep{maoz1995, maoz2005}. In the present work
UV variations are observed in LINERs (HG+13; HG+14). This is consistent with the work
by \cite{maoz2005} at UV frequencies with \emph{HST} data, who showed that UV
variations are common in LINERs. Thus, it seems that LINERs might be showing an
unobstructed view of the inner parts of the AGN. As explained before, this could be
due to the lack of the torus or due to the fact that the torus is not in our line of
sight. In favor of the first scenario, \cite{omaira2015} studied different types of
AGN at mid-infrared frequencies and found that faint LINERs (logL(2--10 keV) = 41 [erg/s]) are consistent with the lack of the torus because the
spectral shape is considerably different to that observed in bright LINERs and
Seyferts and cannot be well reproduced by Clumpytorus models \citep{nenkova2008}
using the BayesClumpy tool \citep{ramos2009}.

X-rays
have the power of penetrating through the torus (whenever they are not Compton-thick) and, because these clouds
are moving very close to the AGN, the clouds show absorption variations registered at
these wavelengths. 
These variations in the absorber have been found in four of our Seyfert 2s, with
location on two of them consistent with the BLR\footnote{Note that for the other two
Seyfert 2s we could not constrain the location of the clouds because the timescales
between observations were too large (HG+14; HG+15).} (HG+15). 
These kind of X-ray eclipses have been observed in Seyferts at different scales from the inner BLR up to the sublimation radius \citep{risaliti2007,puccetti2007,risaliti2010,risaliti2011,braito2013,marinucci2013,markowitz2014}. 
On the contrary, we did not find X-ray eclipses among LINERs, except in NGC\,1052, which seems to behave as a Seyfert 1.8 at X-rays \citep[][HG+14]{omaira2014}. 
An extreme case of absorption variations are the changing-look candidates, where
the source changes from the Compton-thin (Compton-thick) to the Compton-thick
(Compton-thin) regime \citep{guainazzi2002, guainazzi2002a, matt2003b, bianchi2005a}. 
We found that changing-look candidates are only present among Seyfert 2s, while no
LINER changing-look candidates have been reported. 
The fact that X-ray variations related to the absorbers are not detected among LINERs might be suggestive of the lack of the BLR, although variability studies in the appropriate timescales of these sources would be necessary in order to confirm or reject this scenario.

Moreover, the nature of the BLR in LINERs is being debated in the current
literature \citep[][Marquez et al. in prep.]{balmaverde2014, balmaverde2015, constantin2015}. It is being questioned if the broadening of
the balmer lines in LINERs is due to a true BLR or it can be produced by
outflowing material. Only four out of seven LINERs included in the  study
by \cite{constantin2015} do show a true BLR. Thus the LINER 1 nature
needs to be revisited with high spectal and spatial resolution to get a clear
conclussion on the disappearance of BLR at low luminosities.


\section{\label{conclusion}Summary}

In the present work we have assembled the X-ray spectral properties and variability patterns of two samples of AGN selected at optical wavelengths: LINERs and Seyfert 2s. Since 
Compton-thick sources do not usually show variations, the work is centered in Compton-thin sources, including 11 LINERs and 13 Seyfert 2s. 
While the indices of the power law are very similar, slight differences are found in the temperatures, column densities and black hole masses, and major differences are observed in the X-ray intrinsic luminosities and the Eddington ratios.
We have shown that the most frequent X-ray variability pattern occurring between months and years is related with changes in the nuclear continuum in both families, but other patterns of variability are also observed. In particular, variations due to absorbers at hard X-ray energies are most frequent in Seyfert 2s than in LINERs, and variations at soft X-ray energies are rare and need to be confirmed. 

We suggest that the X-ray variations occur similarly in LINERs and Seyfert 2s and might be related with viscous propagating fluctuations in the accretion disc flow, although the accretion mechanisms can be different, being more efficient in Seyfert 2s. Furthermore, based on the scarcity of absorption variations, the lack of changing-look candidates, and the fact that UV nuclear variations are found in these sources, our results are suggestive of at least some LINERs having an unobstructed view of the nucleus, in contrast to its obstructed view in Seyfert 2s. This might be in agreement with theoretical works that predict the torus and/or the BLR disappeareance in at least some LINERs.


\begin{acknowledgements}

          We thank the anonymous referee for his/her helpful suggestions that helped to improve the manuscript. We thank Dr. Del Olmo, Dr. Sulentic, and Dr. Papadakis for their careful reading of the manuscript, and Dr. Netzer for helpful discussions during the work. This work was financed by MINECO grant AYA 2010-15169, AYA 2013-42227-P, and Junta de Andaluc\'{i}a TIC114. LHG acknowledges financial support from the Ministerio de Econom\'{i}a y Competitividad through the Spanish grant FPI BES-2011-043319. This research made use of data obtained from the \emph{Chandra} Data Archive provided by the \emph{Chandra} X-ray Center (CXC). This research made use of data obtained from the \emph{XMM}-Newton Data Archive provided by the \emph{XMM}-Newton Science Archive (XSA). This research made use of the NASA/IPAC extragalactic database (NED), which is operated by the Jet Propulsion Laboratory under contract with the National Aeronautics and Space Administration. We acknowledge the usage of the HyperLeda database (http://leda.univ-lyon1.fr).

\end{acknowledgements}

\bibliographystyle{apj}
\bibliography{LINERSeyfert2}

\appendix

\section{\label{cluster}Clustering analysis}

A clustering analysis was performed in order to build a statistical 
classification of the galaxies in our sample. With this aim we applied 
the scikit-learn sofware package\footnote{http://scikit-learn.org}, a simple and efficient tool for data mining and data analysis based in Machine Learning that 
provides several powerful clustering techniques. Among them we chose the 
Spectral Clustering technique as implemented by \citet[][see also a detailed description by \citealt{Luxburg07atutorial}]{scikit-learn}. 

Spectral Clustering requires the number of clusters or groups to be specified in 
advance. As we will show, the number of selected clusters reflects different physical properties. 
The classification of the groups is built on the basis of $N_{H2}$, $\Gamma$,  L(0.5--2 
keV), L(2--10 keV),  $M_{BH}$, $R_{Edd}$, and a parameter to include whether X-ray variations were detected in the source or not.

A resampling analysis was applied to study how well the
groups are defined using Monte-Carlo simulations. In order to do so, 2000 boostrap realizations of the spectral
clustering were performed to the groups.

Once we selected the number of groups and classified the membership of each nucleus within these groups, we performed a discriminant 
analysis to study which parameters or set of them weight more on the 
classification. Again the scikit-learn software was used. In our case both 
linear ($lda$) and quadratic ($qda$) discriminant analyses were applied to all 
the possible combinations of parameters. 
These discriminants represent two different classifiers, $lda$ is a classifier with a linear decision boundary, whereas $qda$ is a classifier with a quadratic decision boundary. Both are generated by fitting class conditional densities to the data, using a Gaussian density model which is fitted to the parameters.
For each combination of parameters, we 
evaluated the total efficiency as defined by the ratio of correctly classified 
galaxies over the total number. We considered the discriminant parameters as those obtained with the minimum number of parameters and an efficiency larger than 97\% (i.e., 3$\sigma$ of a Gaussian distribution).

\begin{table*}[th]
\caption{Mean properties per group using the Spectral Clustering.}
\begin{center}
\begin{tabular}{lcccccccc}
\hline \hline
\multicolumn{1}{c}{Group}&
\multicolumn{1}{c}{Members}&
\multicolumn{1}{c}{$N_{H1}$}&
\multicolumn{1}{c}{$N_{H2}$}&
\multicolumn{1}{c}{$\Gamma$}&
\multicolumn{1}{c}{L(0.5--2keV)} &
\multicolumn{1}{c}{L(2--10 keV)} &
\multicolumn{1}{c}{$M_{BH}$} &
\multicolumn{1}{c}{$R_{Edd}$ }\\
(1) & (2) & (3) & (4) & (5) & (6) & (7) & (8) & (9) \\ \hline \vspace*{0.1cm}
LINER A  & NGC\,315, NGC\,1052, NGC\,1961,  & 0.01$^+_-$0.01 & 9.1$^+_-$5.3  &   1.7$^+_-$0.2  &  41.2$^+_-$0.2  &   41.3$^+_-$0.2  &  8.6$^+_-$0.3  &   -4.4$^+_-$0.5  \\
\null &  NGC\,4261 \\
\hline
LINER B & NGC\,2681 NGC\,4278 NGC\,4374,  &  0.06$^+_-$0.11  &  1.1$^+_-$2.9  &  1.8$^+_-$0.2  &  39.7$^+_-$0.7  &  39.8$^+_-$0.6  &  7.9$^+_-$0.6  &  -5.4$^+_-$0.8  \\
\null &  NGC\,3718, NGC\,4494, NGC\,4552, \\
\null & NGC\,4736, NGC\,5195, NGC\,4698,\\
\null &    NGC\,5982  \\
\hline
Seyfert 2 A & MARK\,348, NGC\,788, ESO\,417-G01, &  0.13$^+_-$0.26  & 26.1$^+_-$14.0  &  1.6$^+_-$0.4  &  42.2$^+_-$0.5  &  42.7$^+_-$0.4  &  7.5$^+_-$063  &  -1.7$^+_-$0.6  \\
\null & 3C98.0, MARK\,1210, NGC\,4507,   \\
\null & MARK\,268, MARK\,273,  MARK\,477,  \\
\null & IC4518A, ESO\,138-GG01, NGC\,6300,   \\ 
\null &  NGC\,7172, NGC\,7319 \\
\hline
Seyfert 2 B & NGC\,424, MARK\,573, MARK\,1066,  &   0.01$^+_-$0.04  &  46.9$^+_-$19.1  &  0.5$^+_-$0.2  &  41.1$^+_-$0.9  &  41.4$^+_-$0.9  &  7.4$^+_-$0.7  &  -3.2$^+_-$0.8  \\
\null &  MARK\,3, IC\,2560, NGC\,3393,  \\
\null & NGC\,5194, Circinus, NGC\,5643, \\
\null &  NGC\,7212 \\
\hline
\end{tabular} \\
\end{center} 
{{\bf Notes.} (Col. 1) Groups, (Col. 2) membership of galaxies, (Cols. 3--9) mean and standard deviation of the spectral parameters for each group.}
\label{tab:gmean}
\end{table*}

A caveat of this analysis might be that our sample does not include the lower luminosity Seyfert 2s (see Sect. \ref{previous}). At the end of the section we specify the implications of the inclusion of these sources.

When Spectral Clustering is applied with just two clusters to define the 
partition, the sample is split into two groups that correspond exactly to 
LINERs and Seyfert 2s, with the exception of NGC\,4698 which is labelled as being 
a LINER (see Sect. \ref{spectra}). The discriminant analysis shows that the parameter which has more influence in the separation of the groups is $R_{Edd}$, but the efficiency of this classification is 89\%; to obtain a 97\% of confidence level the parameters responsible for these groups are $R_{Edd}$ and $\Gamma$. 

When we perform the clustering with three groups, Seyfert 2s are distributed 
in two groups which correspond to the Compton-thin (i.e., Seyfert 2 A) and Compton-thick (i.e., Seyfert 2 B) Seyfert 2
subtypes. The clustering fails only in classifying MARK\,477 and ESO\,138-G01, which we classified as Compton-thick candidates (HG+15) and are included in the Seyfert 2 A group. The parameters responsible for these groups are again $R_{Edd}$ and $\Gamma$ with a 97\% of confidence level, since only one parameter ($R_{Edd}$) results in an efficiency of lower than 90\%. The clustering also shows that Seyfert 2 B includes non-variable sources.

When the clustering is performed with four groups, LINERs are also 
distributed in two subgroups, which are different in their X-ray luminosities. 
We checked whether this classification differentiate between type 1.9 and 2 LINERs, and found that it does not. Two type 1.9 and another two type 2 LINERs are included in LINERs A, and three type 1.9 and six type 2 LINERs are included in LINERs B\footnote{We did not take into account NGC\,4698 for the counting.}.
The previous split of Seyfert 2 galaxies is kept. 

When five groups are used for the clustering, it splits the Seyfert 2 B sample in two, with one of the groups including only two sources (IC\,2560 and NGC\,5643). But no clear differences are observed in the spectral parameters between these two groups. 

The resampling analysis was therefore applied to the four groups. We performed 2000 boostrap realizations of the spectral
clustering with these groups and for each of them, every
galaxy was assigned to the nearest cluster. We find that the two Seyfert 2 groups are really stable and also
the LINER B group. However, this is not the case of the LINER A group. 
All the galaxies in this group moved to the LINER B group in almost half of the
simulations. By contrast, the galaxies of the other groups remain in
the original sampling more than the 80\% of the cases. We
decided to keep separated the four LINER A galaxies since they may show
different spectral properties. The conclusion on the existence of these two
LINER populations should wait the analysis of a larger sample.

Both the $qda$ and $lda$ analyses indicate that if we perform the discriminant 
analysis with only one parameter, the one that provides the highest efficiency (i.e., the 
strongest discriminator) is $R_{Edd}$, but it is impossible to perform the 
classification with it since the resulting efficiency amounts only to 74\%. 
According to the $qda$, just two parameters ($\Gamma$ and L(2-10~keV)) would be 
enough for performing a full clustering whereas according to the $lda$, to 
obtain a larger efficiency than 97\%, at least three parameters are needed. 
Both $lda$ and $qda$ agree on that with the combination of one of the luminosities, 
$\Gamma$, and $R_{Edd}$, it is possible to obtain the same grouping 
than with all the parameters.

\begin{figure*}
\centering
\includegraphics[width=0.81\linewidth]{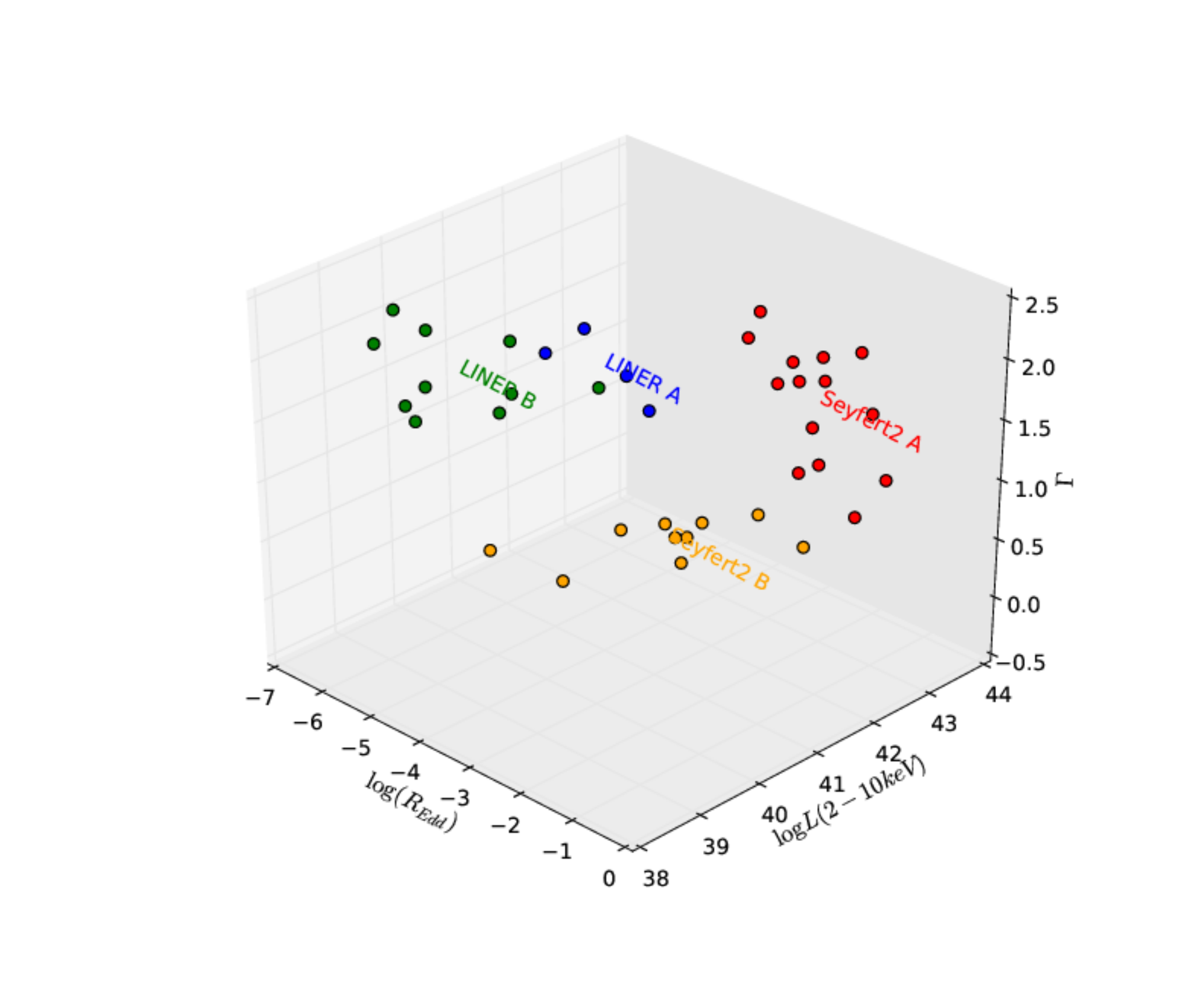}

\includegraphics[width=0.33\linewidth]{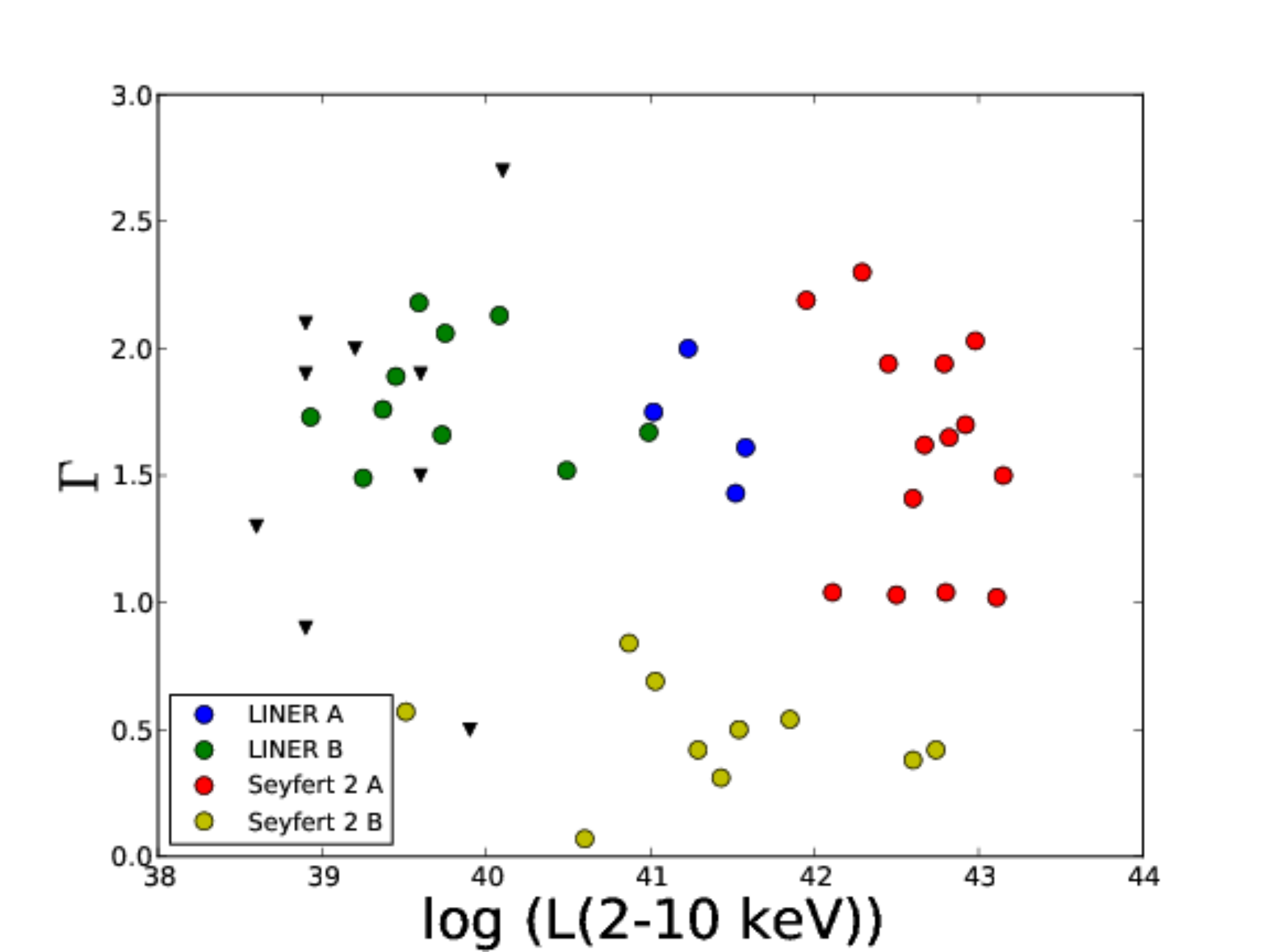}
\includegraphics[width=0.33\linewidth]{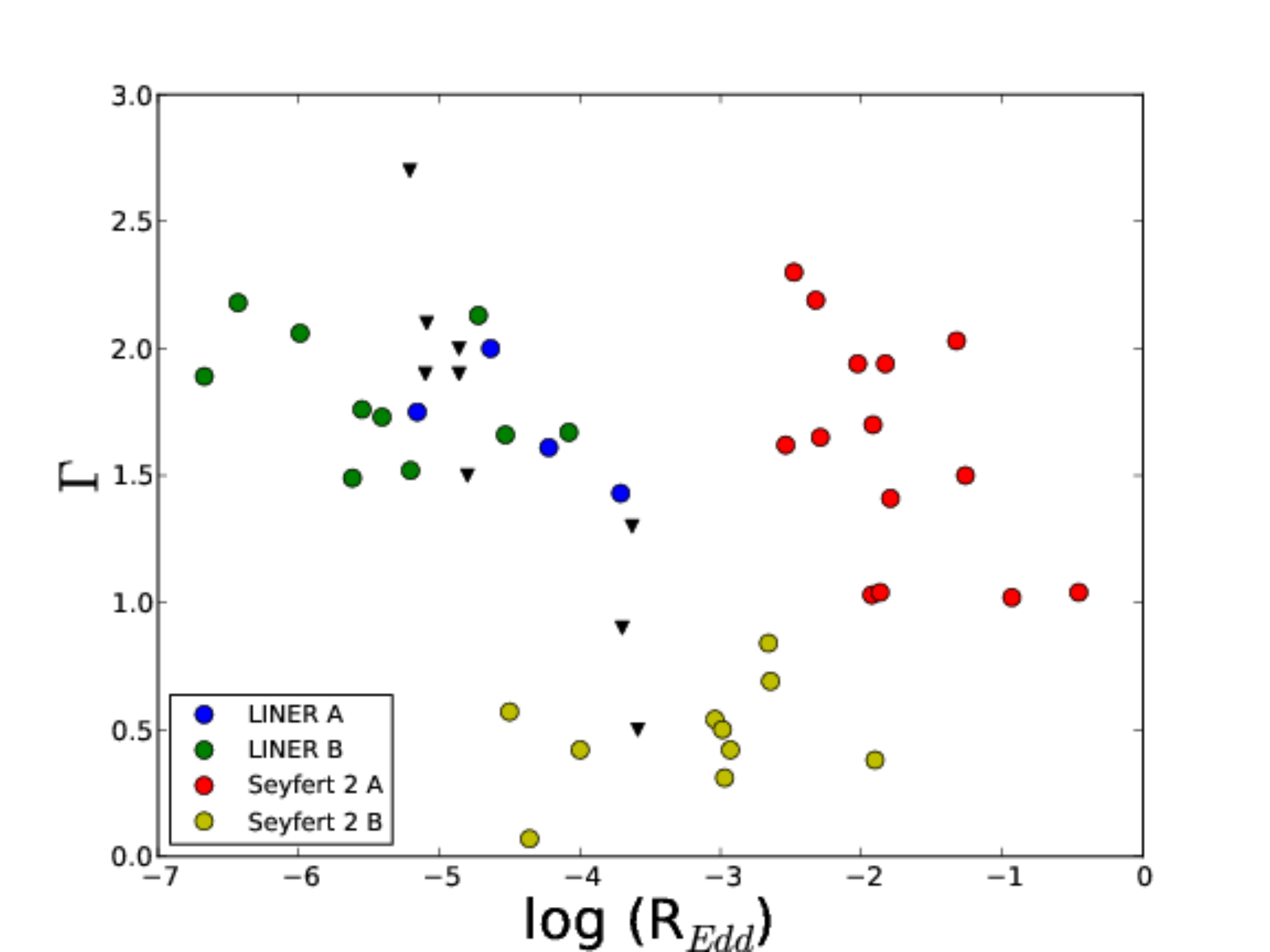}
\includegraphics[width=0.33\linewidth]{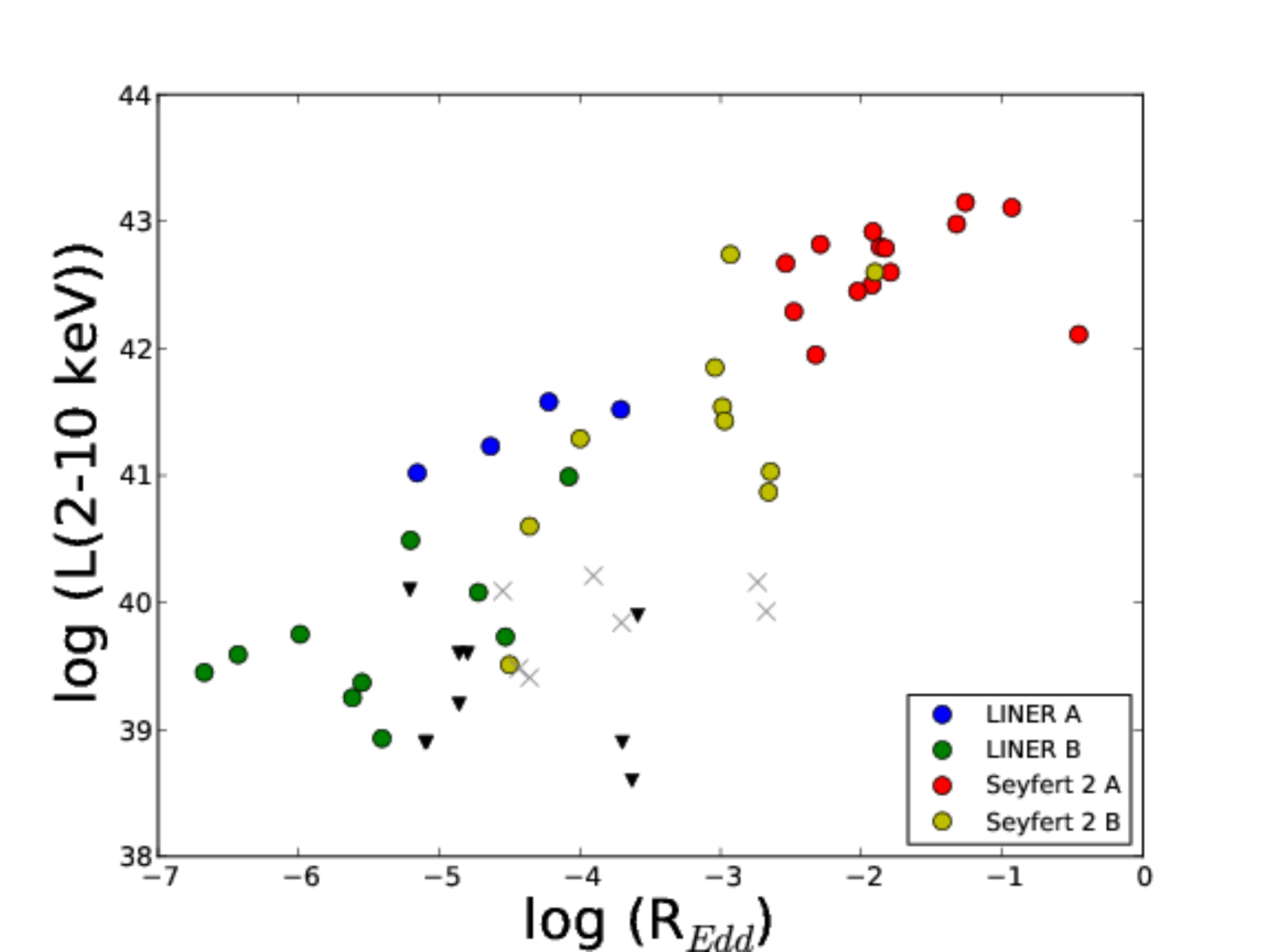}
\caption{Spectral Clustering types in the ($R_{Edd}$, L(2--10 keV), 
$\Gamma$) space. Lower panels show the 2D projections of the three parameters space. See details on the group membership in Table \ref{tab:gmean}. Blue circles represent the LINER A group, green circles the LINER B group, red circles the Seyfert 2 A group, yellow circles the Seyfert 2 B group, black triangles the low luminosity Compton-thin Seyfert 2s from \cite{cappi2006}, and gray crosses the low luminosity Compton-thin Seyfert 2s from \cite{brightman2011}. We estimated $M_{BH}$ (in order to calculate $R_{Edd}$ as in \citealt{eracleous2010}) from the low luminosity Seyfert 2s from the velocity dispersion given in Hyperleda. }
\label{gdibu}
\end{figure*}

Our main results are summarized in Table \ref{tab:gmean}, where the membership 
for each group and the mean 
parameters per group together with their standard deviations are presented.
A representation of the grouping included in the ($R_{Edd}$, L(2--10 keV), 
$\Gamma$) space of parameters is presented 
in Fig. \ref{gdibu}, along with the projections for a proper visualization.
It has to be 
noticed
that in our selected sample we missed the low luminosity Seyfert 2s, which
exist in the \cite{brightman2011} and \cite{cappi2006} samples. In Fig. \ref{gdibu} the low 
luminosity
data from \cite{cappi2006} have been included (black triangles) in the three projections and the \cite{brightman2011}
only in the logL(2--10 keV) vs $R_{edd}$ plot since for these data $\Gamma$ have been fixed 
to 1.9.
Thereof it appears that low luminosity Seyfert 2s in all plots are mixed
with LINER B group.

The clustering analysis shows that LINERs and Seyfert 2s cannot be differentiated by the observed X-ray variability. Since Seyfert 2s are distributed by their Compton-thick and Compton-thin classifications, it is able to classify sources that do not show variations in the Seyfert 2 B group, but the Seyfert 2 A group includes also sources where variations were not detected. LINERs A and B groups include one (out of four) and seven (out of ten) variable sources.

\end{document}